\documentclass[]{pasj02}
\usepackage{natbib}
\usepackage{url}
\usepackage{subcaption}

\jyear{2025}
\Received{}
\Accepted{}

\begin{document}

\title{Coronal Non-Thermal and Doppler Plasma Flows Driven by Photospheric Flux in 28 Active Regions}

\author{
    James \textsc{McKevitt},\altaffilmark{1,2}\altemailmark\orcid{0000-0002-4071-5727}\email{james.mckevitt.21@ucl.ac.uk}
    Sarah \textsc{Matthews},\altaffilmark{1}\orcid{0000-0001-9346-8179}
    Deborah \textsc{Baker},\altaffilmark{1}\orcid{0000-0002-0665-2355}
    Hamish A. S. \textsc{Reid},\altaffilmark{1}\orcid{0000-0002-6287-3494}
    David H. \textsc{Brooks},\altaffilmark{3,1}\orcid{0000-0002-2189-9313}
    Ignacio \textsc{Ugarte-Urra},\altaffilmark{4}\orcid{0000-0001-5503-0491}
    Peter R. \textsc{Young},\altaffilmark{5,6}\orcid{0000-0001-9034-2925}
    Teodora \textsc{Mihailescu},\altaffilmark{7}\orcid{0000-0001-8055-0472}
}

\altaffiltext{1}{University College London, Mullard Space Science Laboratory, Holmbury St Mary, Dorking, Surrey, RH5 6NT, UK}
\altaffiltext{2}{University of Vienna, Institute of Astrophysics, Türkenschanzstrasse 17, Vienna A-1180, Austria}
\altaffiltext{3}{Computational Physics, Inc., Springfield, VA 22151, USA}
\altaffiltext{4}{Space Science Division, Naval Research Laboratory, Washington, DC 20375, USA}
\altaffiltext{5}{NASA Goddard Space Flight Center, Solar Physics Laboratory, Heliophysics Science Division, Greenbelt, MD 20771, USA}
\altaffiltext{6}{Department of Mathematics, Physics and Electrical Engineering, Northumbria University, Newcastle upon Tyne, UK}
\altaffiltext{7}{INAF Osservatorio Astronomico di Roma, Monte Porzio Catone 00078, Italy}


\maketitle

\begin{abstract}
Magnetohydrodynamic (MHD) waves and/or the braiding of magnetic field lines are largely thought to be responsible for heating the solar corona, both being mechanisms which are driven by the Sun's photospheric magnetic field. Recent modelling work leads us to expect that such heating mechanisms would be seen in the excess broadening (non-thermal velocity) of coronal spectral emission lines and that larger magnitudes of photospheric magnetic flux would generate more heating, but a direct connection between magnetic flux and spectral line broadening has been difficult to establish. We combine measurements of the photospheric magnetic field from SDO/HMI and non-thermal velocity in log~$T$$\sim$6.2 coronal plasma from Hinode/EIS for 28 active regions and find a moderate correlation between the two exists in quiescent active regions, consistent with the photospheric field injecting upward Poynting flux into the solar corona and causing coronal heating. We find that no strong correlation with coronal composition makes it difficult to distinguish between MHD wave heating and magnetic field braiding heating using these diagnostics with current instrumentation.
\end{abstract}


\section{Introduction}

The cause of the solar corona's extreme million-degree temperature remains a fundamental open problem in solar physics. Two broad classes of models are thought to heat coronal plasma: nanoflare or reconnection heating, in which countless small-scale magnetic reconnection events dissipate stored magnetic energy into heat \citep[e.g.,][]{parker_nanoflares_1988}, and wave or turbulence heating, in which magnetohydrodynamic (MHD) waves (e.g. Alfvén or kink waves) launched by convective motions propagate upward and dissipate their energy in the corona \citep[e.g.,][]{alfven_magneto_1947,ionson_resonant_1978}. In both scenarios, the source of energy is the Sun's magnetic field and convective motions at the photosphere, which inject Poynting flux into the corona. In the nanoflaring scenario, footpoint shuffling can tangle (braid) coronal field lines and generate numerous current sheets, releasing energy in many small reconnection bursts. Similarly in the wave heating scenario, twisting motions or vortex flows at the photosphere can drive MHD waves which, through non-linear wave-wave interactions, cascade to smaller scales in a turbulent spectrum and at small enough scales dissipate as heat. Through these processes, photospheric motions effectively act as an energy pump, injecting upward Poynting flux that can balance the enormous radiative and conductive losses of active region plasma. Estimates by \cite{parker_magnetic_1983} and subsequent modelling by e.g. \cite{yeates_coronal_2014} suggest that the energy input from braiding could be sufficient to explain the coronal conductive and radiative losses in active regions. Furthermore, several observations by for instance \cite{viall_evidence_2012,ishikawa_detection_2017} find evidence consistent with the nanoflare heating model \citep{klimchuk_key_2015}. There have been suggestions that coronal heat originates in the chromosphere \citep[e.g.,][]{aschwanden_coronal_2007, de_pontieu_observing_2009}, but subsequent work finds differently \citep[e.g.,][]{klimchuk_are_2014,patsourakos_core_2014}. Meanwhile \cite{mcintosh_alfvenic_2011} have found evidence for Alfvénic waves sufficient to heat the quiet corona. However, Alfvén speeds sufficient to explain coronal heating throughout the real density-stratified corona have not yet been observed \citep{van_doorsselaere_coronal_2020,morton_alfvenic_2023}.

Both these nanoflare and MHD wave heating mechanisms rely on injection of energy into the corona from the photosphere, quantified by the vertical Poynting flux. It is known that this Poynting flux is proportional to the photospheric magnetic flux, with \cite{welsch_photospheric_2015} finding on the order of 10$^5$~erg/s of energy are injected into the corona per Mx at the photosphere. This implies that the photospheric magnetic flux of an active region is intimately linked to the energy input available for heating and plasma motions. Indeed, \cite{pevtsov_relationship_2003} presented results suggesting such a relationship in all active stars.

Spectroscopic observations of coronal plasma, required to develop our understanding of these mechanisms, have been primarily performed by the extreme ultraviolet (EUV) imaging spectrometer \citep[EIS;][]{culhane_euv_2007} onboard the Hinode spacecraft \citep{kosugi_hinode_2007} over the last nearly 20~years. Spectral emission lines are typically observed by Hinode/EIS as quasi-Gaussian distributions, which are also sometimes considered as $\kappa$-distributions or similar with large wings (see interesting discussion in \cite{lorincik_plasma_2020}). Such functions are fitted to the observed spectral lines, with their widths representing the thermal broadening of the plasma, broadening due to the point spread function of the instrument, and then additional Doppler broadening which has previously been attributed to a variety of processes \citep{del_zanna_solar_2018}.

These processes include unresolved Alfvénic motions generated by convective footpoint driving \citep{hassler_line_1990, banerjee_signatures_2009}, turbulence induced by impulsive heating events such as nanoflares \citep{patsourakos_nonthermal_2006}, and the superposition of small-scale flows that remain unresolved within the instrumental resolution \citep{hara_coronal_2008}. Other proposed contributions include line-of-sight integration effects across multiple structures \citep{feldman_unresolved_1983}, and broadening produced by interchange reconnection and associated rarefaction waves in coronal outflow regions \citep{bradshaw_reconnection-driven_2011}. The stochastic dynamics of spicules and related jet-like features (causing asymmetric broadening) have also been suggested \citep{de_pontieu_observing_2009}, as has non-equilibrium ionisation, particularly in rapidly evolving transition region and coronal plasma \citep{bradshaw_self-consistent_2003, dzifcakova_diagnostics_2011}. However, neither spicule- or non-equilibrium ionisation-related heating are currently thought to make a notable contribution to excess broadening in quiescent coronal plasma around 1~MK.

Where this excess broadening is expressed as an equivalent velocity (non-thermal velocity), previous simulation work has found this to be correlated with magnetic braiding-induced turbulence \citep{pontin_non-thermal_2020} and Alfvén wave turbulence \citep{asgari-targhi_observations_2024}. In the case of braiding-type heating, plasma is thought to be initially heated to flare-like temperatures on the order of log~$T$$\sim$7 and then cool. A study by \cite{winebarger_can_2004} found that in the case that active region loops are impulsively heated, temperature and density measurements at lower temperatures are insufficient to recover information about what may have caused their previous heating. However, the aforementioned modelling by \cite{pontin_non-thermal_2020} produced braiding-induced broadening at the cooler coronal log~$T$$\sim$6.2, raising the question of whether broadening might be a parameter that retains some information about previous heating.

As coronal spectral line broadening may be observed where coronal heating has occurred, that coronal heating is a product of Poynting flux, and that Poynting flux is increased above stronger photospheric magnetic fields, we might then see a relationship between this photospheric magnetic field strength and excess spectral line broadening in the corona. However, \cite{brooks_measurements_2016} reported no significant trend between the excess broadening in high temperature active region loops and the active region's unsigned photospheric magnetic field strength when considering 15 active regions. They concluded that the limited broadening we observe in the cooling phase of coronal loops is all that remains well after the energy release process is complete. Their study was focused on active region loop measurements, but observations of different features or plasma including different structures across a whole active region might yield a different result. It would also then be interesting to consider the evolution of an active region's excess broadening with age as when its photospheric magnetic field disperses, it could be expected that the excess broadening would decrease either because the region's magnetic stressing and nanoflaring subside and/or because plasma conditions become more uniform.

As both Alfvén waves and nanoflares should cause excess spectral line broadening, an additional spectral diagnostic is needed to distinguish these mechanisms observationally, and here we turn to composition measurements.

The composition of coronal plasma can be determined using the First Ionization Potential (FIP) effect, which describes the enhancement or depletion of elemental abundances in the solar corona relative to their photospheric values \citep{laming_fip_2015}. Elements with a relatively low-FIP are typically enhanced in abundance in the corona compared to the photosphere, whereas high-FIP elements tend to maintain their photospheric abundance in the corona. This enhancement is quantified using the FIP bias parameter which is the ratio of the coronal elemental abundance to the photospheric elemental abundance. One leading theory explaining the FIP effect describes it as driven by the ponderomotive force associated with MHD waves, like Alfvén waves, propagating through or reflecting from the chromosphere. The ponderomotive force acts selectively on chromospheric ions but not on neutral atoms, leading to elemental fractionation \citep{laming_non-wkb_2012}. This effect is observed using Hinode/EIS by comparing the intensities of low-FIP elements (typically Si) and high-FIP elements (like S), additionally taking into account temperature- and density-related effects \citep[e.g.,][]{brooks_full-sun_2015}. A previous study by \cite{baker_plasma_2013} found a moderate correlation between the non-thermal velocity and FIP bias of coronal plasma in the footpoints of an active region, supportive of the ponderomotive force model and potentially indicative of Alfvén waves. On larger scales, however, the relationship is unexplored, and complicated by the fact there are competing views on the location and driver of the MHD waves associated with the elemental fractionation \citep{laming_first_2017,martinez-sykora_impact_2023}.

In this study, we investigate the relationships between coronal non-thermal velocity and various other active region properties in 28 active regions captured by full-Sun Hinode/EIS spectral scans. Comparable studies typically rely on small samples of dedicated active region observations, which introduce a bias toward compact regions or fail to capture entire structures given the limited Hinode/EIS field of view. The full-disk approach (combining observations to cover the full Sun) captures an unbiased variety of active regions, including extended structures with very different morphologies that would not otherwise be studied. We combine this Hinode/EIS data with photospheric magnetic flux measured by the Helioseismic Magnetic Imager \citep[HMI;][]{scherrer_helioseismic_2012} onboard the Solar Dynamics Observatory \citep[SDO;][]{pesnell_solar_2012}, active region type and age as determined by \cite{mihailescu_what_2022}, and coronal composition as calculated by \cite{brooks_full-sun_2015}, to address: 1) if the modelled proportionality between magnetic flux and upward Poynting flux translates to statistically-significant observed heating-induced line broadening on active region scales, 2) if FIP bias and line broadening measurements can help distinguish between heating mechanisms, 3) if any observed heating signatures depend on active region age and evolutionary stage. We make the code to automatically construct Hinode/EIS full disks available\footnote{This paper uses version 0.1.2: \url{https://doi.org/10.5281/zenodo.17641183}. The source code and ongoing development are available at: \url{https://github.com/jamesmckevitt/eismaps}}.

\section{Observations}

\begin{figure*}
    \centering
    \includegraphics[width=\textwidth]{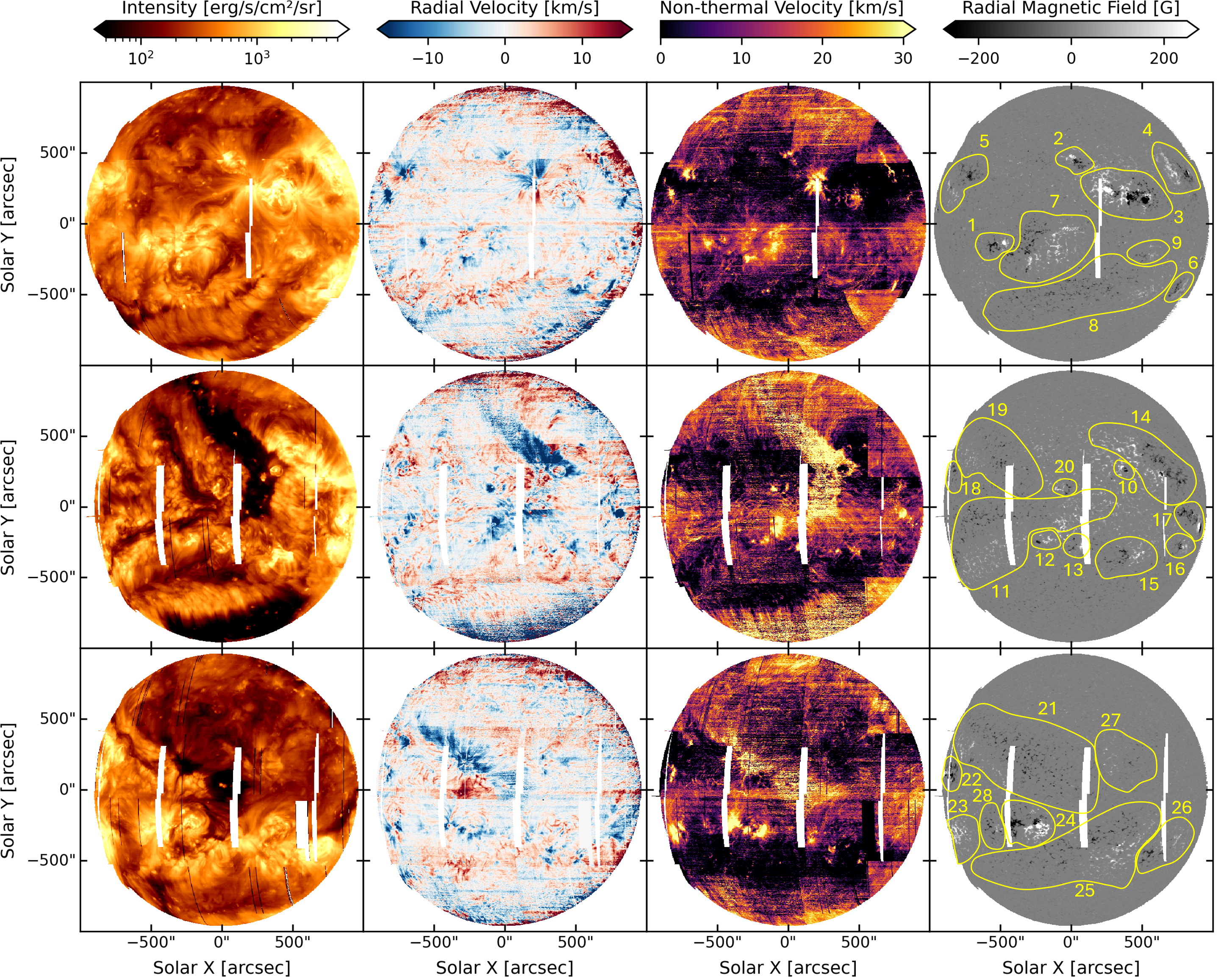}
    \caption{Full Sun maps for 16-18th January 2013 (top row), 1-3 April 2015 (middle row), 18-20 October 2015 (bottom row) showing from Hinode/EIS the Fe~XII~195.119 intensity (far left column), Fe~XII~195.119 radial velocity (centre left column), Fe~XII~195.119 non-thermal velocity (centre right column), and from SDO/HMI the radial photospheric magnetic field component (far right column). The active regions considered in this study are highlighted with yellow contours and numbered in the right-hand column.\\
    Alt text: Four measurements of full-disk solar maps from three observing periods, each showing intensity, Doppler velocity, non-thermal velocity, and radial magnetic field. Active regions are outlined and numbered.}
    \label{fig:full_disks}
\end{figure*}

\subsection{EUV Observations}

We use three full-disk mosaic data sets taken by the scanning slit EUV spectrometer Hinode/EIS between (times in UT) 09:37 on 16 January 2013 and 06:00 on 18th January 2013 (disk 1), 09:14 on 1 April 2015 and 00:45 on 3 April 2015 (disk 2), and 10:27 on 18 October 2015 and 00:26 on 20 October 2015 (disk 3). During these three windows, Hinode/EIS was operating low-cadence, wide field-of-view active region studies, positioned to cover the whole solar disk in 26 separate rasters.

These rasters are positioned so that together they cover the whole solar disk, where some overlap between the rasters is present. Due to the differential rotation of the solar surface beneath the spacecraft, and the fact the rasters to cover one full disk take just less than two days to complete, the rasters should be adjusted to positions correct relative to one another, and overlap regions handled with care.

We use the flux-conserving spherical polygon intersection method from \texttt{Astropy} \citep{the_astropy_collaboration_astropy_2022} to position our rasters appropriately relative to the first raster of each full disk. Regions where multiple rasters covered the same region on the Sun will be measured differently by the spectrometer, partly due to instrumental noise effects (see \cite{mckevitt_pre-flare_2026}) and partly because in the sometimes many hours between rasters, plasma in that region will have changed. In the coronal intensity (Fe~XII~195.119~\AA{} log~$T\sim6.2$) full disk maps seen in Figure~\ref{fig:full_disks} we take the raster contributing the highest intensity to each pixel.

We note that the rasters of the first full disk were ordered differently to the other two we consider here, which is why the white stripes (indicating no data is available for that part of the disk) appear different (EIS studies 491 and 544 were used)\footnote{\url{https://solarb.mssl.ucl.ac.uk/SolarB/}}. These studies complete a scanning raster, with 123 pointing positions taken sequentially from west to east with a 2~arcsec slit and with a scan step size of 4~arcsec, meaning the active regions were recorded with quasi-4~arcsec spatial resolution.

\subsubsection{Bulk flows}

To measure coronal plasma motions we used the strong Fe~XII~195.119~\AA{} (log~$T\sim6.2$) coronal line. This line has the highest signal-to-noise ratio of any coronal lines observed by Hinode/EIS \citep{young_euv_2007}, and as the full disk mosaics include coverage of many quiet Sun areas with low signal it provided the lowest $\chi^2$ residual when fitting the observed spectra.

We perform a two-component Gaussian fitting for each pixel in each raster scan using \texttt{MPFit} \citep{markwardt_non-linear_2009}, as implemented in the \texttt{EISPAC} code \citep{weberg_eispac_2023}, with a seperate component for the Fe~XII~195.119~\AA{} and Fe~XII~195.179~\AA{} blended lines respectively \citep{del_zanna_benchmarking_2005}, where a fixed 0.06~\AA{} offset in the components is imposed. Taking the fit parameters of the Fe~XII~195.119~\AA{} component, we calculate the line of sight Doppler velocity using 

\begin{equation}
    v_{\text{Doppler}} = \frac{c(\lambda_{\text{obs}}-\lambda_{0})}{\lambda_{0}},
\end{equation}

\noindent{}where $c$ is the speed of light, $\lambda_{\text{obs}}$ is the observed wavelength at peak intensity of the fit, and $\lambda_{0}$ is the rest wavelength of the line. We found no pixels to be saturated during these observations and so did not exclude any for this reason from fitting. The accuracy of Doppler velocity measured by Hinode/EIS is known to be approximately 5~km/s (\citealt{culhane_euv_2007,young_velocity_2012}; see also \citealt{kamio_modeling_2010} for an interesting discussion on instrument uncertainties.)

As we consider active regions at different positions on the disk and there is known to be some dependence on the line-of-sight velocity with disk position \citep{demoulin_3d_2013}, we normalise for this by adjusting each pixel's measured line-of-sight Doppler velocity for its position on the solar disk using the equation

\begin{equation}
    v_{\text{norm}} = \frac{v_{\text{Doppler}}}{\cos{\theta}},
\end{equation}

\noindent{}where $\theta$ is the angle between the line of sight and the normal to the solar surface. This is a simplification which assumes any Doppler velocities measured across the disk are observations made at an angle to plasma flowing normal to the surface. The alternative approach would be to avoid the adjustment, where instead measurements of Doppler velocity are washed out towards the limb, making a direct comparison between active regions at different points on the disk difficult. We, therefore, apply this adjustment in our analysis. A consequence of this approach is the amplification of measurement uncertainties near the limb where, as $\cos{\theta}$ becomes small, the effective noise is increased. This broadens the distribution of plasma measured with near-zero velocities. However, as seen in our results later, this affects only active regions at the very edge of the disk and is seen only in lower pixel counts in the narrow range of $|v_\text{norm}| \lesssim 1$~km/s.

\subsubsection{Non-thermal motions}

We convert the width of the fitted Gaussian to a non-thermal velocity ($v_{nt}$), that being the measured width of the observed spectral lines unexplained by the thermal and instrumental widths expressed as a velocity, using 

\begin{equation}
    {\text{FWHM}_{o}}^2={\text{FWHM}_{i}}^2+4\ln{2}\left(\frac{\lambda}{c}\right)^2\left({v_t}^2+{v_{nt}}^2\right),
\end{equation}

\noindent{}where $\text{FWHM}_{o}$ and $\text{FWHM}_{i}$ refer to the observed and instrumental full width at half maximum values respectively, and where $\lambda$ and $v_t$ refer to the emission line rest wavelength and the associated thermal velocity respectively. We use the instrumental width along the slit determined by \cite{young_instrumental_2011}, and the thermal widths from \texttt{eis\_width2velocity} in the EIS software tree in Solarsoft.

The uncertainty in the instrumental width can be combined with the statistical error in the fitted Gaussian~-~partly caused by uncertainty in the measurement of each point in the emission line~-~using standard error propagation \citep[e.g.,][]{bevington_data_2003} to estimate the error in non-thermal velocity measurements to be approximately 25\%. Filtering pixels by their $\chi^2$ value was attempted to reduce noise but found not to improve the resulting maps. No significant center-to-limb variation is seen in excess spectral line broadening \citep[e.g.,][]{chae_sumer_1998}, and so no adjustment to these line widths based on position on disk is necessary.

\subsubsection{Plasma composition}

To probe coronal abundance enhancements we used Si~X~258.38~\AA{} (FIP=8.15~eV) and S~X~264.23~\AA{} (FIP=10.36~eV) emission. We calculate the FIP bias, where

\begin{equation}
    \text{FIP}_{\text{bias}}=\frac{\text{coronal elemental abundance}}{\text{photospheric elemental abundance}},
\end{equation}

\noindent{}using the intensities of the relatively low-FIP Si~X~258.38~\AA{} and relatively high-FIP S~X~264.23~\AA{} emission lines. This combination is appropriate for studying log~$T\sim6.2$, ideal for the quiescent active regions covered in this study and corresponding with the log~$T\sim6.2$ plasma motions we investigate. The FIP values are those presented by \cite{mihailescu_what_2022}, where the method used is that described in \cite{brooks_establishing_2011} and \cite{brooks_full-sun_2015} being designed to remove density effects with the 
Fe~XIII~202.04~\AA\ to 203.82~\AA\ ratio \citep{watanabe_fe_2009, young_euv_2007}, and temperature effects using the Fe lines {\small\rmfamily VIII} to {\small\rmfamily XVI} to derive the differential emission measure (DEM). These intensities were calibrated using the in-flight radiometric calibration curves provided by \cite{warren_absolute_2014}. We note a new calibration is available \citep{del_zanna_hinode_2025}, but since the calibrations only differ after the observation dates considered here, and to maintain consistency with \cite{mihailescu_what_2022}, we do not apply the new calibration.

The pointing information of the Hinode/EIS data was corrected by co-aligning the Fe~XII~195.12~\AA\ intensity maps with imaging performed at 193~\AA\ by the Atmospheric Imaging Assembly \citep[AIA;][]{lemen_atmospheric_2012} onboard SDO. This required only small corrections to both longitude and latitude.

\subsection{Magnetic Field Observations}

We also used data gathered by SDO/HMI at each of the raster times of Hinode/EIS. SDO/HMI generates full-disk vector photospheric magnetograms with a cadence of 12~minutes and at a resolution of about 1~arcsec, with a per-pixel noise level ($\sigma_{noise}$) of about 100~Gauss \citep{hoeksema_helioseismic_2014}. However, we calculate properties of active regions using many pixels ($N$). Since the standard error of sample quantiles (including the median and percentiles we use here) scales as $\sigma_{noise}/\sqrt{N}$ \citep{bevington_data_2003}, this spatial averaging reduces the statistical uncertainty to the order of 1~G for each active region. Here we consider the radial component $B_r$ of these 12~minute vector magnetograms as opposed to the higher-cadence line-of-sight magnetograms so as to use a measure of magnetic field strength normalised to position on the disk, given we are comparing active regions at various disk positions. We acknowledge that photospheric magnetic flux measurements from HMI are increasingly affected by noise towards the limb \citep{hoeksema_helioseismic_2014}. However, as we are looking at active regions with strong polarities and we only find this affect very barely noticeable in one active region, we find the measurements acceptable for our purposes.

\subsection{Active Regions}

\begin{table}[]
    \centering
        \begin{tabular}
        {lp{.9cm}p{1.6cm}|lp{.9cm}p{1.6cm}}
        \hline
        \hline
        AR & Age (days) & Ev. Stage        & AR & Age (days) & Ev. Stage        \\ \hline
        1         & 5   & Spot             & 15        & 24  & Decayed          \\
        2         & 11  & Decayed          & 16        & 11  & Decayed          \\
        3         & 11  & Spot             & 17        & 12  & Spot             \\
        4         & 13  & Spot             & 18        & 3   & Decayed          \\
        5         & 39  & Dispersed        & 19        & 58  & Filament channel \\
        6         & 13  & Decayed          & 20        & 7   & Decayed          \\
        7         & 53  & Dispersed        & 21        & 29  & Filament channel \\
        8         & 189 & Filament channel & 22        & 30  & Spot             \\
        9         & 13  & Decayed          & 23        & 27  & Spot             \\
        10        & 0.5 & Emerging             & 24        & 27  & Spot             \\
        11        & 120 & Filament channel & 25        & 63  & Dispersed        \\
        12        & 6   & Decayed          & 26        & 38  & Dispersed        \\
        13        & 8   & Decayed          & 27        & 35  & Dispersed        \\
        14        & 8   & Dispersed        & 28        & 21  & Decayed \\
        \hline
        \end{tabular}
    \caption{Ages and evolutionary stages (Ev. Stage) of active regions (AR) considered in this study, as defined by \cite{mihailescu_what_2022}.}
    \label{tab:active_region_ages}
\end{table}

The active regions in this study are those identified and analysed by \cite{mihailescu_what_2022}. They were defined by eye using SDO/HMI magnetograms, where boundaries are drawn around concentrations of magnetic flux such that contours are broad enough to include all magnetic flux associated with the active region, but where pixels with a magnetic flux density below 30~G are filtered out of the analysis to avoid considering small-scale background field between the active region field fragments. In a similar fashion, two polarities within each active region are identified. In the region for a given polarity, no opposite polarity field is considered in any analysis. We use the evolutionary stages and ages as identified and defined by \cite{mihailescu_what_2022}, summarised in Table~\ref{tab:active_region_ages}. Following the \lq{}emerging\rq{} phase, they define the \lq{}spot\rq{} phase where bipoles are still clear (peak or early decay), \lq{}decayed\rq{} (sunspots have disappeared), \lq{}dispersed\rq{} (active region photospheric field barely distinguishable from quiet Sun), and active regions with filament channels \citep{mihailescu_what_2022}.

We note that the emerging active region (10) appears as an outlier in our results, displaying high non-thermal velocity values. It can be seen in Figure~\ref{fig:full_disks} to reside in a coronal hole where stray light in Hinode/EIS is known to artificially increase measured intensity \citep{young_scattered_2022}, affecting non-thermal velocity measurements. Furthermore, it is the smallest active region in our study by area, and so our active region-scale parameters are averaged over less pixels. We, therefore, exclude it from any statistical trends we draw from the data analysis.

\section{Results}

\subsection{Metrics}

\begin{figure}

  \begin{subfigure}{0.48\textwidth}
    \includegraphics[width=\linewidth]{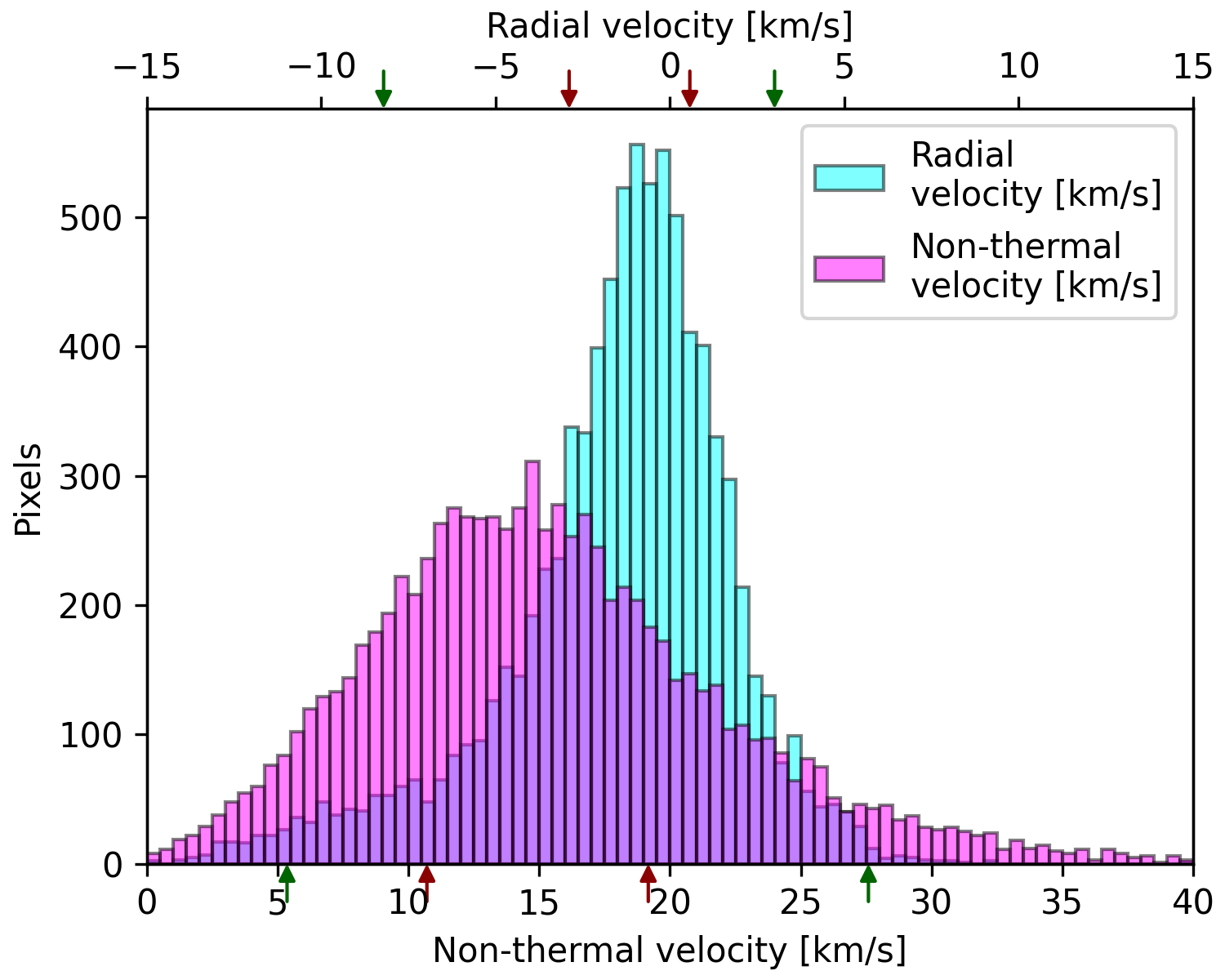}
    \caption{Distribution of non-thermal velocity and radial velocity (derived from Doppler velocity) in active region 1.}
    \label{fig:ntv_vel_hist_1}
  \end{subfigure}

  \hfill

  \begin{subfigure}{0.48\textwidth}
    \includegraphics[width=\linewidth]{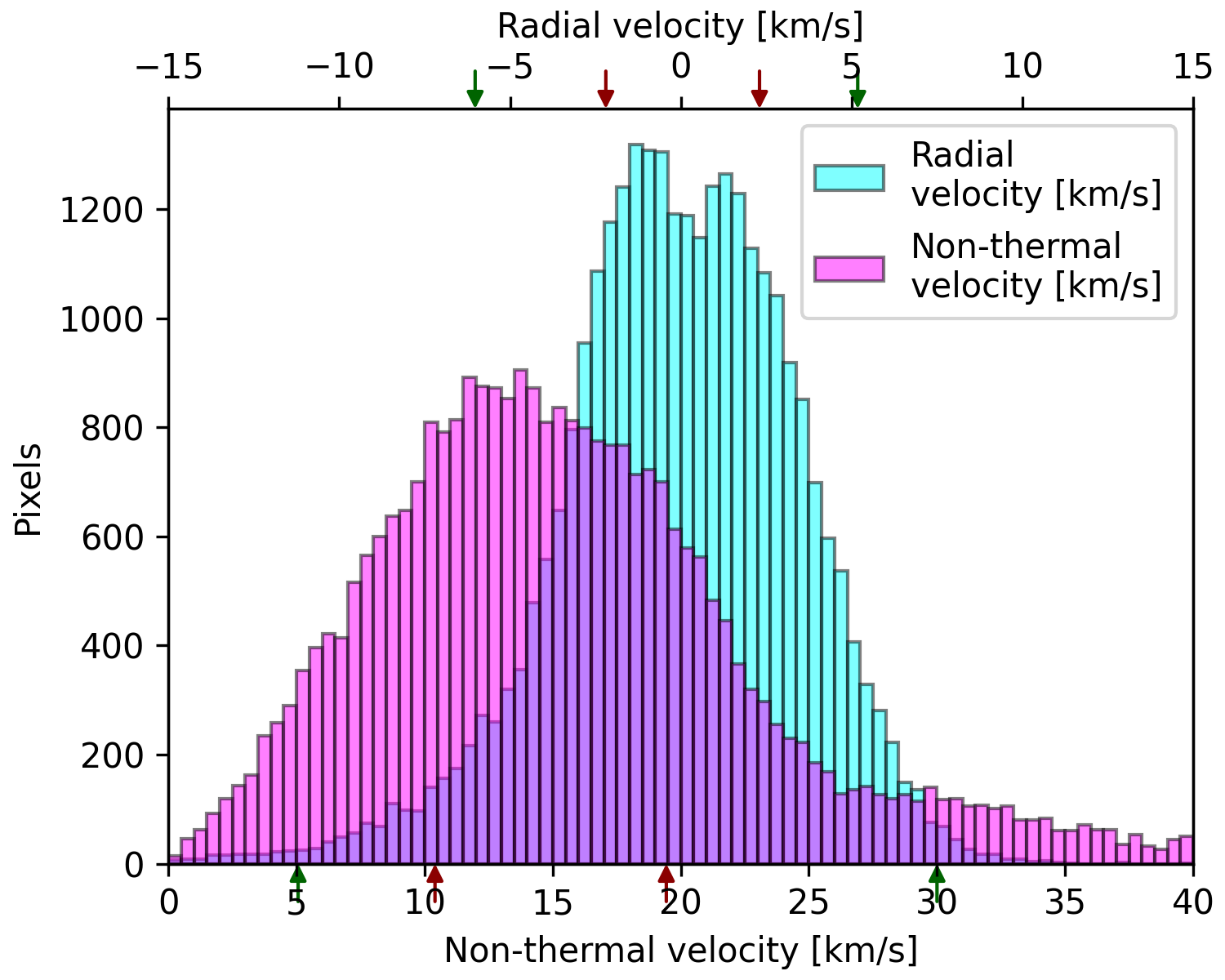}
    \caption{Distribution of non-thermal velocity and radial velocity (derived from Doppler velocity) in active region 23.}
    \label{fig:ntv_vel_hist_23}
  \end{subfigure}

  \vspace{1\baselineskip}

  \caption{Histograms of non-thermal velocity and radial velocity (derived from Doppler velocity) measurements in active regions 1 and 23. The 25th and 75th percentiles are shown with red arrows and the 5th and 95th percentiles with green arrows, on the top and bottom axes for their respective measurements.\\
  Alt text: Histograms for two active regions showing distributions of non-thermal velocity and radial velocity. Percentile markers indicate differences in spread.}
  \label{fig:ntv_vel_hists}

\end{figure}

To quantitatively explore the relationships between coronal velocity and photospheric magnetic field we employ a range of metrics, such as the Pearson product-moment correlation coefficient (hereinafter referred to as $r$) by \cite{bravais_analyse_1846}. We also use the interquartile range (IQR) - that being the difference between the 75th and 25th percentiles - and the 90\% interpercentile range (90\%~IPR) - that being the difference between the 95th and 5th percentiles - to quantify the spread of the bulk of the data, and to quantify a range excluding outlying pixels respectively. 

Such population distributions of pixels within active regions can be seen in Figure \ref{fig:ntv_vel_hists}. In both the top and bottom panels, for active regions 1 and 23 respectively, we see the velocity distributions have an asymmetric tail towards upflowing velocities. In the top panel (a) we see this somewhat represented in the IQR but much more so in the 90\%~IPR. In the bottom panel (b), where the active region is larger and encompasses more pixels, we see the tail is not able to make any noticeable impact on the IQR, but is noticeable in the 90\%~IPR.

In the case of the non-thermal velocity values, both active regions share a similar distribution peaking at between 10 and 15~km/s before tailing off. Active region 23 (panel b) shows some small secondary component peaking at 30~km/s, extending the tail of the distribution for this active region slightly. We see that the IQR is unaffected by this as it is very similar for both active regions, but the 90\%~IPR reflects this in its upper bound.

We present these distributions and percentiles of two representative active regions to inform the results we show below for the complete dataset, as we use both the IQR and the 90\%~IPR.

\subsection{Non-thermal velocity versus Magnetic Field}

\begin{figure}
    \centering
    \includegraphics[width=\columnwidth]{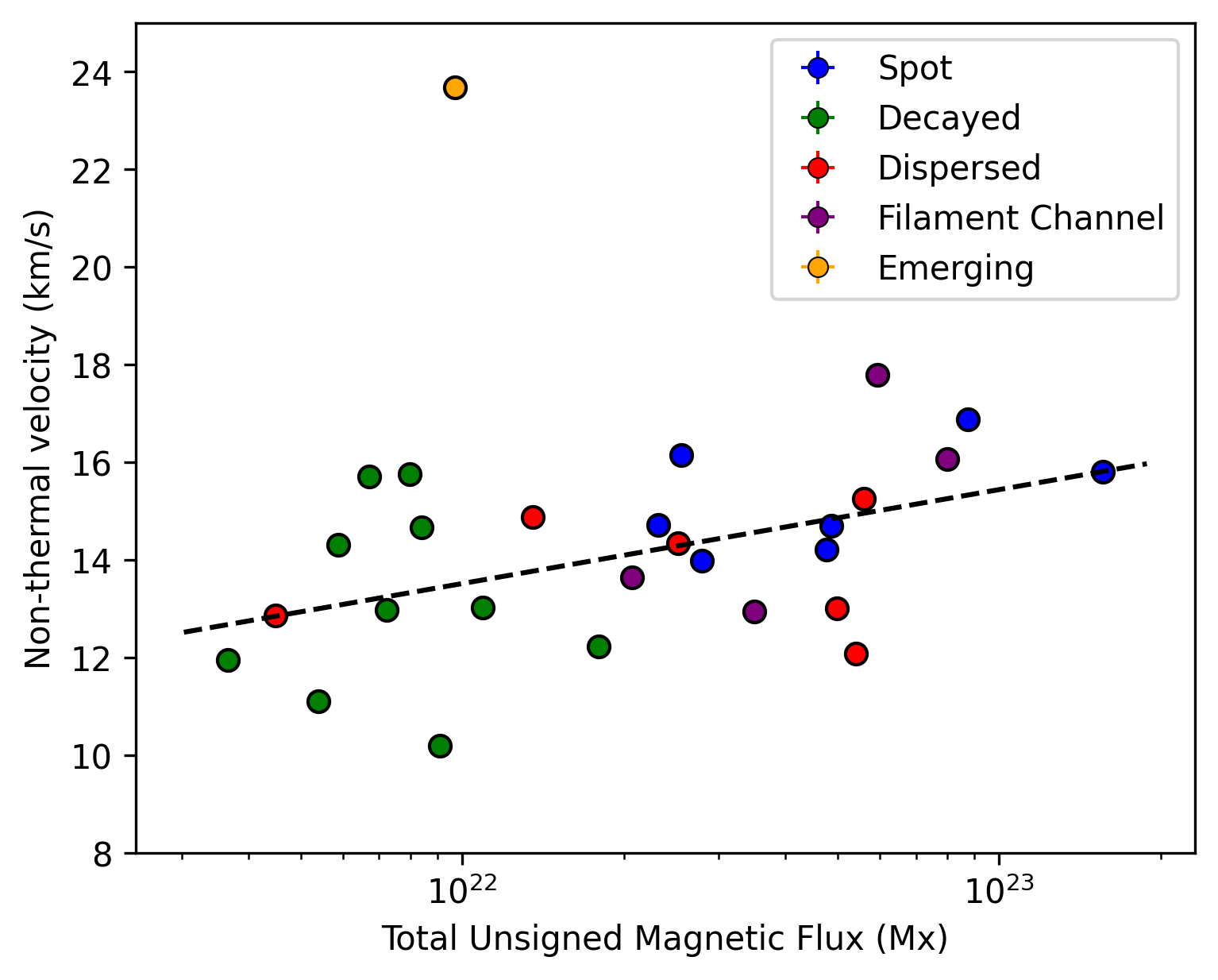}
    \caption{Total radial unsigned magnetic flux ($\sum{}|B_r|\cdot{}A$) against the median non-thermal velocity for each active region, where $A$ is the pixel area in cm$^2$.\\
    Alt text: Scatter plot comparing non-thermal velocity with total magnetic flux.}
    \label{fig:ntv-mag}
\end{figure}

Figure~\ref{fig:ntv-mag} shows the relationship between non-thermal velocity and the total radial photospheric magnetic flux in Mx in the active regions considered in this study. We see that the decayed active regions, with lower values of total unsigned magnetic flux tend to have slightly lower median non-thermal velocities, whereas the spot type active regions, with the higher total unsigned magnetic flux, appear to have higher median non-thermal velocity values. When including the filament and dispersed-type active regions as well, but excluding the outlier emerging-type active region, we find a moderate positive correlation of $r$=0.48 when using log-transformed total unsigned magnetic flux. We found a similar correlation between log total unsigned magnetic flux and the 90th percentile non-thermal velocity values.

\begin{figure}
    \centering
    \resizebox{\hsize}{!}{\includegraphics{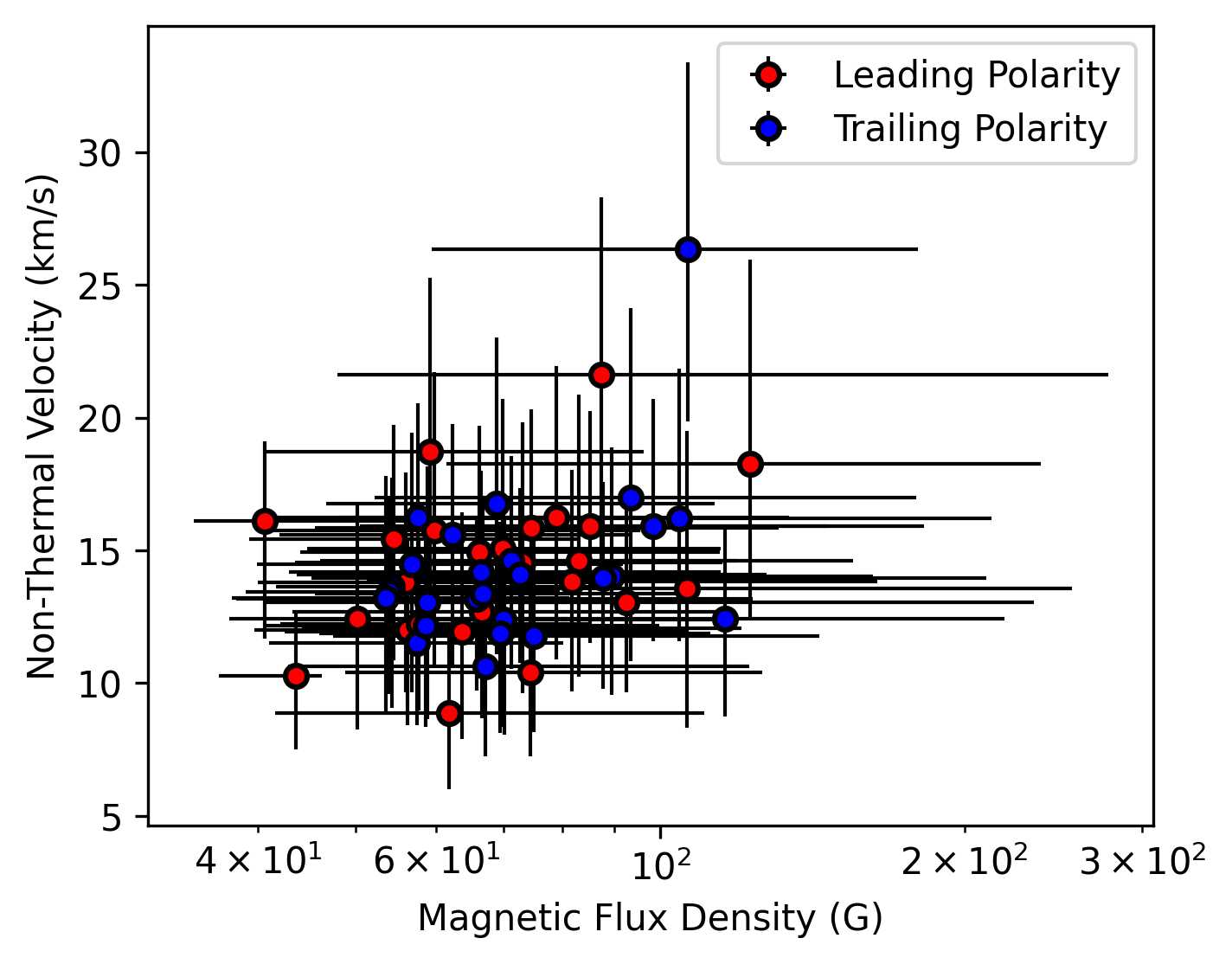}}
    \caption{The relationship between the non-thermal velocity and unsigned magnetic flux density of the leading and trailing polarities of the active regions, with the 25th and 75th percentiles shown with bars and median values with spots.\\
    Alt text: Scatter plot of non-thermal velocity versus unsigned magnetic-flux density for leading and trailing polarities. Median points and percentile ranges are shown, illustrating similar behaviour in both polarities and increasing non-thermal velocity with stronger fields.}
    \label{fig:pol_ntv-magden}
\end{figure}

We also consider such relationships for the leading and trailing polarities of the active regions in this study, shown in Figure~\ref{fig:pol_ntv-magden}. No clear distinction between the behaviour of the polarities is seen here, with both the leading and trailing polarities showing the general trend of above, that higher median non-thermal velocity values are moderately associated with higher median unsigned magnetic flux densities in the active regions.

\subsection{Non-thermal velocity versus Age}

\begin{figure}
  \centering
  \resizebox{\hsize}{!}{\includegraphics{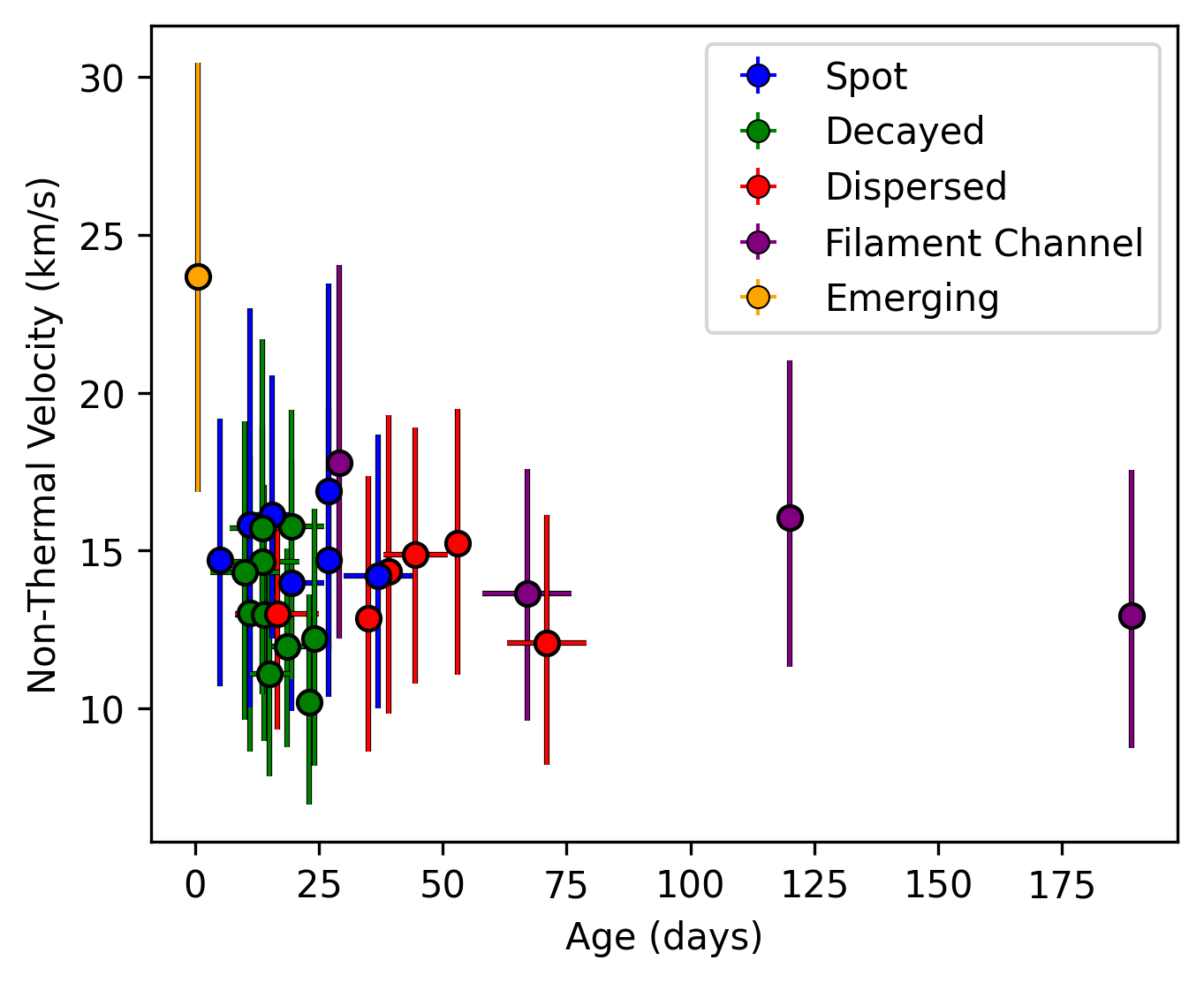}}
  \caption{The relationship between non-thermal velocity and age of the active regions. The median values are shown with circles, and the spread of non-thermal velocities between the 25th and 75th percentiles is shown with bars. For ages, the bars indicate the ranges taken from \cite{mihailescu_what_2022}.\\
  Alt text: Scatter plot of non-thermal velocity versus active-region age, with median markers and percentile ranges. Data show wide variability at all ages and no clear age-dependent trend.}
  \label{fig:ntv-age}
\end{figure}

We see no clear correlation between the non-thermal velocity and the age of the active regions, as shown in Figure \ref{fig:ntv-age}, excluding the emerging active region as an outlier. We similarly observe no clear trend in the spread of non-thermal velocity with age. A consideration of evolutionary stage may be more appropriate for future studies, as we discuss in Section~\ref{sec:discussion}.

\subsection{Non-thermal velocity versus Doppler Velocity}

We also consider the relationship between the distribution of the plasma's non-thermal velocity and radial velocity, derived from the line-of-sight Doppler velocity, for each active region.

\begin{figure}
  \centering
  \resizebox{\hsize}{!}{\includegraphics{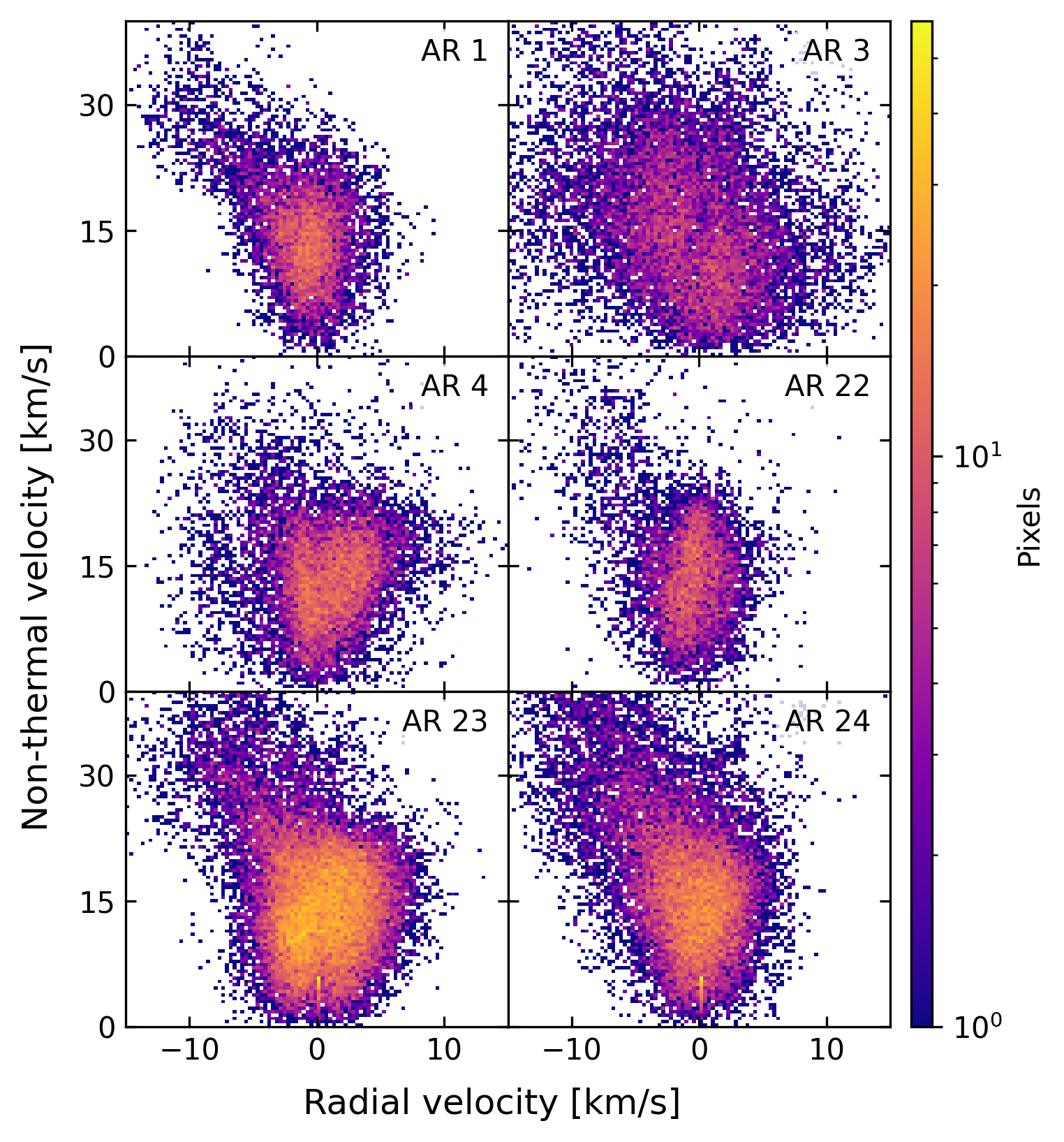}}
  \caption{Histograms of non-thermal velocity against radial velocity (derived from Doppler velocity) for several spot active regions.\\
  Alt text: Six two-dimensional histograms for spot-type active regions showing non-thermal velocity versus radial velocity. Higher non-thermal velocities correspond to increasingly blue-shifted plasma and broader velocity spreads.}
  \label{fig:vel_ntv_hist_spo}
\end{figure}

For several of the spot-type active regions, shown in Figure~\ref{fig:vel_ntv_hist_spo}, we see that most plasma with higher non-thermal broadening is also seen to be upflowing where, as non-thermal velocity increases, the upflow velocities increase. The plasma distributions all broadly increase their spread of radial velocities for higher non-thermal velocity values, up to around $V_{nt}\sim$20~km/s and a spread of $V_r\sim$15~km/s, where the spread decreases and the average flow direction for a given non-thermal velocity becomes increasingly blue-shifted as non-thermal velocity increases. We note active region 4 in the middle left panel doesn't show this relationship as strongly, and displays some increasing red shift to downflowing plasma at $V_{nt}\sim$20~km/s. However, this active region is an outlier amongst all the spot-type active regions, and its plasma distribution is sampled from fewer pixels than, for example, active regions 23 and 24, where the increasing tendancy to upflow with higher non-thermal velocities is more prominent.

\begin{figure}
  \centering
  \resizebox{\hsize}{!}{\includegraphics{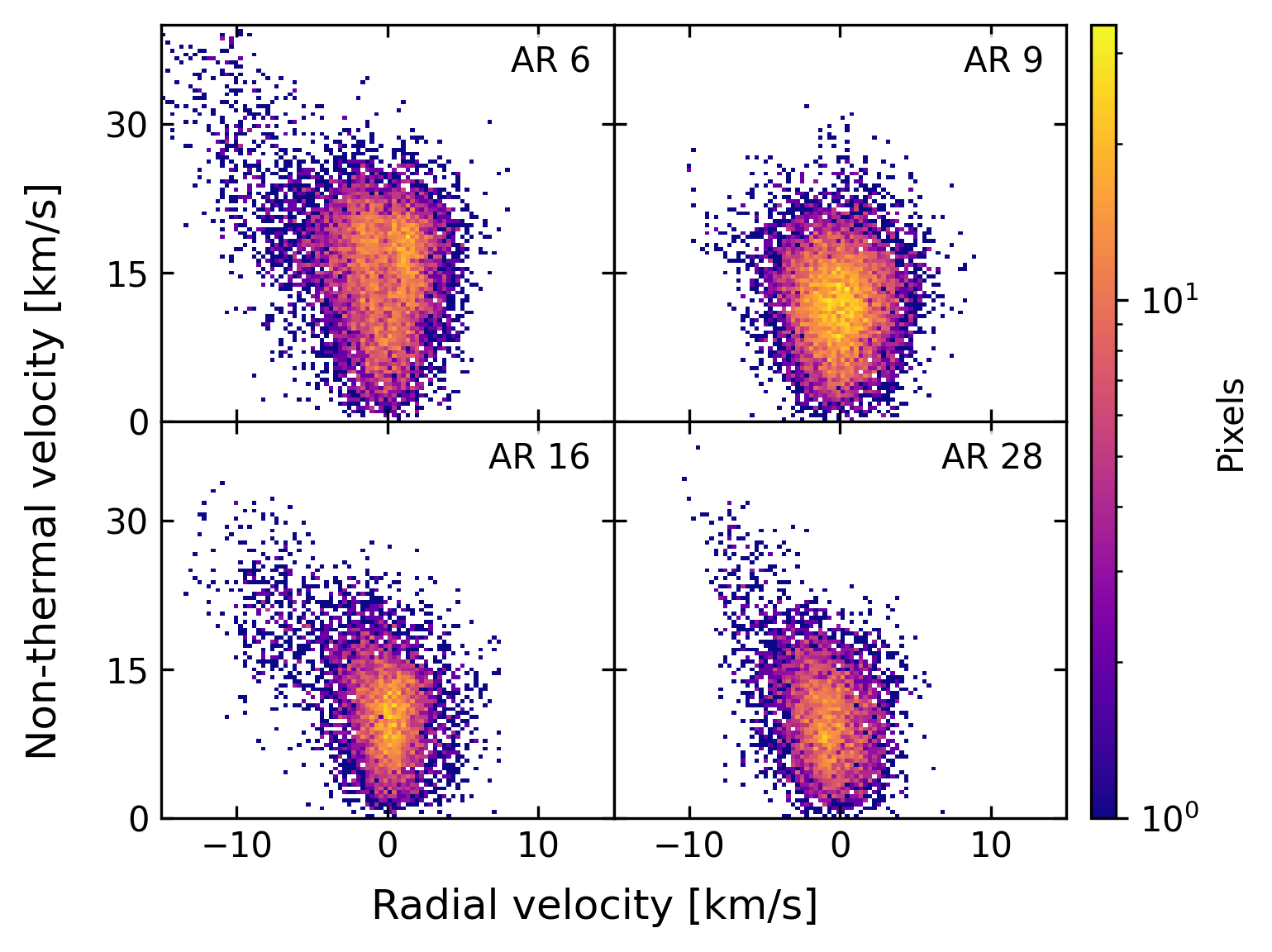}}
  \caption{Histograms of non-thermal velocity against radial velocity (derived from Doppler velocity) for several decayed active regions.\\
  Alt text: Four two-dimensional histograms for decayed active regions showing non-thermal velocity versus radial velocity. Moderate upflow trends appear at lower non-thermal velocities, with overall narrower velocity ranges.}
  \label{fig:vel_ntv_hist_dec}
\end{figure}

We show a similar relationship but for some of the decayed active regions in Figure~\ref{fig:vel_ntv_hist_dec}. These active regions are smaller in area than the spot-type active regions of Figure~\ref{fig:vel_ntv_hist_spo} and so represent lower statistical certainty in plasma distribution. However, we still see the general trend of increasing upflow with higher non-thermal velocities. The transition to average upflowing plasma for a given non-thermal velocity seems to occur at slightly lower non-thermal velocities than for spot-type active regions, where active regions 6, 16 and 28 all begin their upflow skew just below 15~km/s. These decayed active regions can also be seen to display a lower maximum spread in radial velocity values than for spot-type active regions, where almost all plasma is contained within at most a spread of $V_r\sim$10~km/s for a given non-thermal velocity for each active region. We note active region 9 does not appear to display an average radial velocity bias at higher non-thermal velocity values.

\begin{figure}
  \centering
  \resizebox{\hsize}{!}{\includegraphics{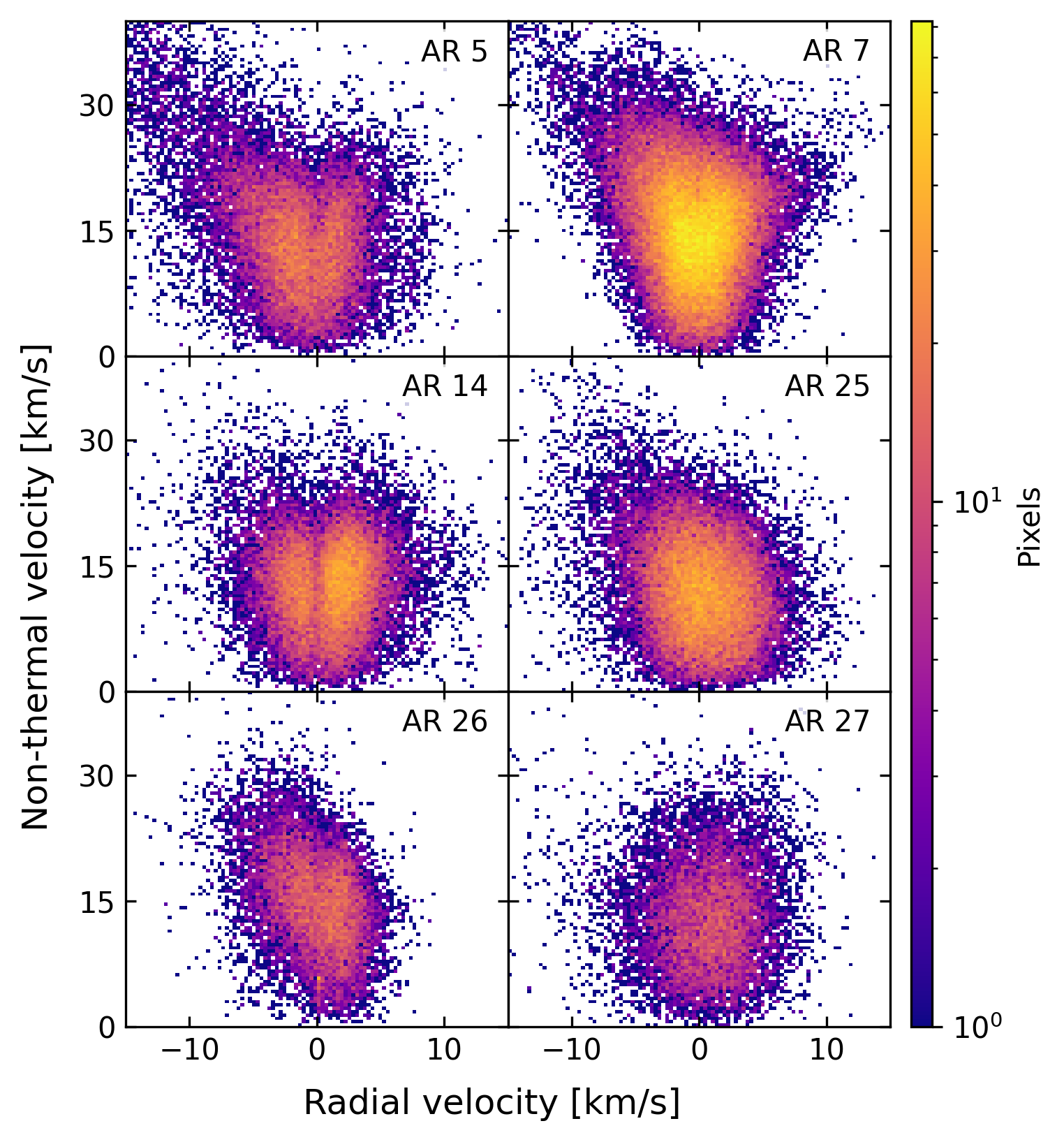}}
  \caption{Histograms of non-thermal velocity against radial velocity (derived from Doppler velocity) for the dispersed active regions.\\
  Alt text: Six two-dimensional histograms for dispersed active regions comparing non-thermal and radial velocities. Some regions show weak upflow trends at higher non-thermal velocities, while others show broad symmetric velocity spreads.}
  \label{fig:vel_ntv_hist_dis}
\end{figure}

\begin{figure}
  \centering
  \resizebox{\hsize}{!}{\includegraphics{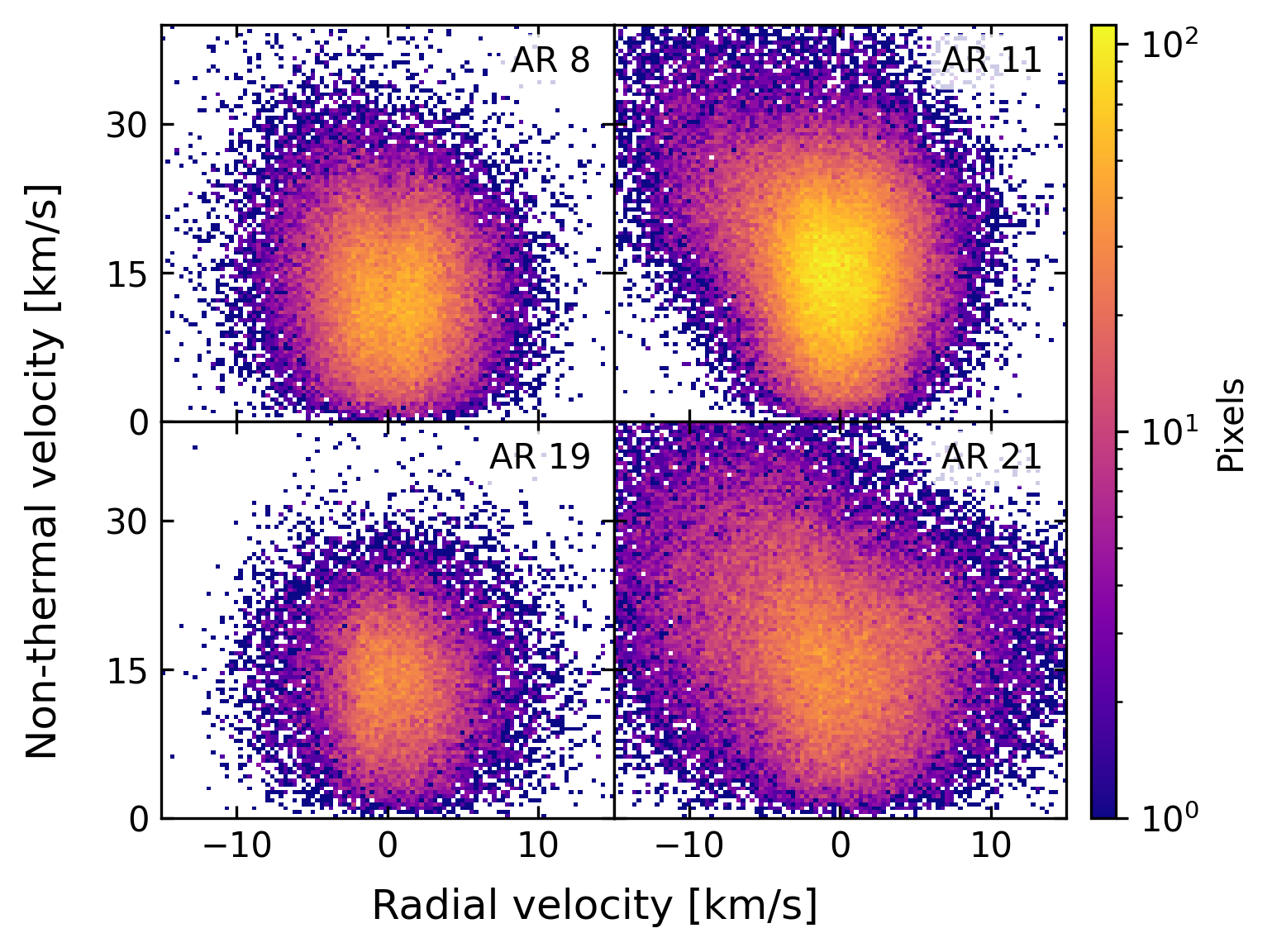}}
  \caption{Histograms of non-thermal velocity against radial velocity (derived from Doppler velocity) for the filament channel active regions.\\
  Alt text: Four two-dimensional histograms for filament-channel active regions showing non-thermal velocity versus radial velocity. Most regions exhibit wide velocity distributions with subtle or absent upflow trends.}
  \label{fig:vel_ntv_hist_fil}
\end{figure}

We consider the distribution of plasma in a similar way for both the dispersed and filament channel-type active regions in Figure~\ref{fig:vel_ntv_hist_dis} and Figure~\ref{fig:vel_ntv_hist_fil} respectively. In such cases, we are typically sampling many more pixels of plasma as such active region types are larger, as seen by the adjusted colour bars compared with Figures~\ref{fig:vel_ntv_hist_spo} and~\ref{fig:vel_ntv_hist_dec}. We see in Figure~\ref{fig:vel_ntv_hist_dis} and Figure~\ref{fig:vel_ntv_hist_fil} that while prominent in some active regions, these dispersed and filament channel-type active regions display a lower tendency to have upflowing plasma at higher values of non-thermal velocity. From the dispersed active regions, active regions 5, 7 and 26 display a clear trend towards upflowing plasma at higher non-thermal velocity values. This trend is loosely apparent in active region 25 but not apparent in active regions 14 or 27. From the filament channel-type active regions, active regions 11 and 21 show some trend towards upflowing plasma at higher non-thermal velocities, but this is subtle compared to the much wider spread of their radial velocities at all values of non-thermal velocity, and not apparent in active regions 8 or 19.

\begin{figure}
  \centering
  \resizebox{\hsize}{!}{\includegraphics{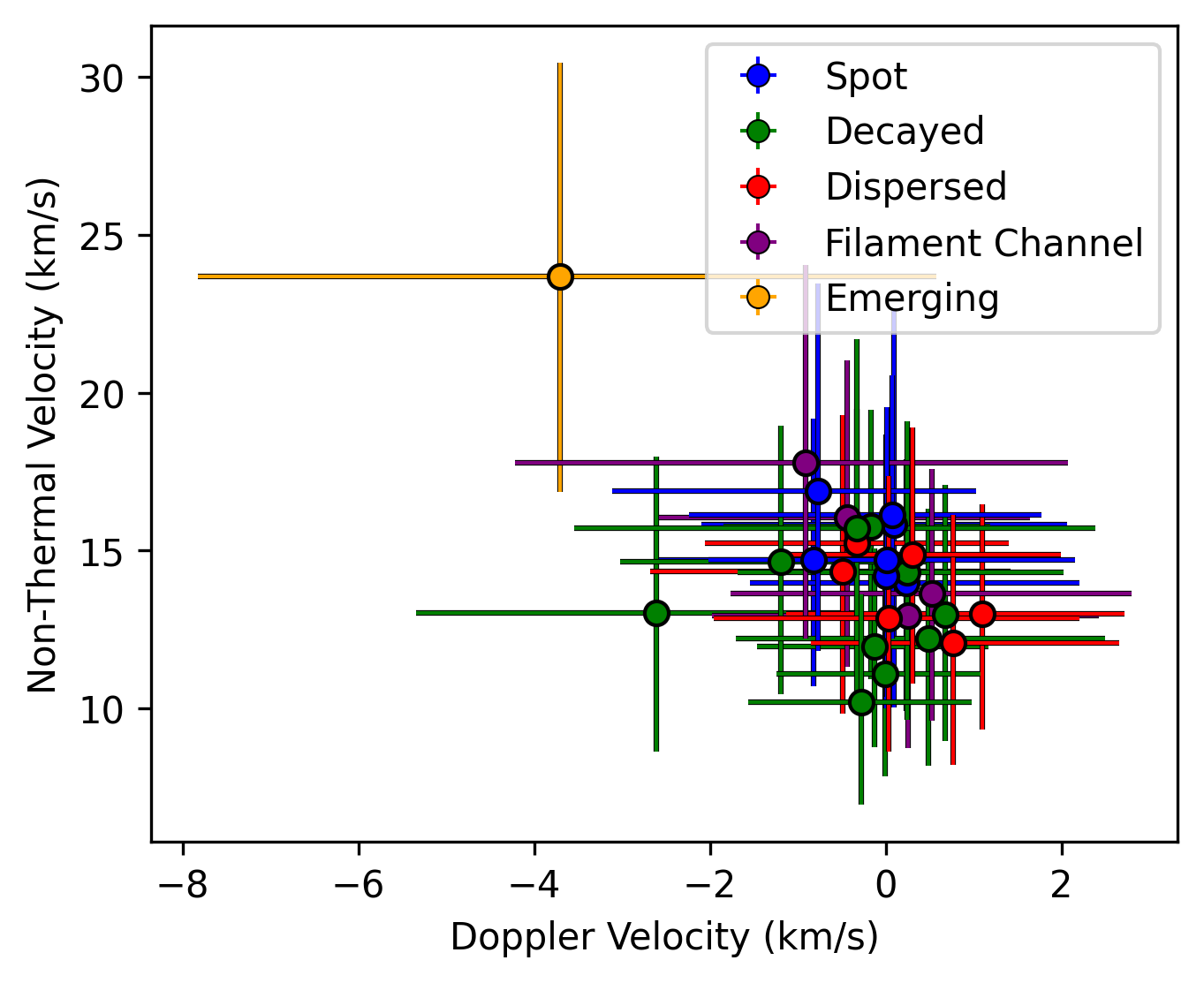}}
  \caption{Scatter plot showing the relationship between non-thermal velocity and radial velocity (derived from Doppler velocity) for all active regions. The median values are shown with circles, and the spread between the 25th and 75th percentiles is also indicated.\\
  Alt text: Scatter plot of median non-thermal velocity versus median radial velocity for all active regions, with percentile bars. Regions with higher non-thermal velocity increasingly show net upflows.}
  \label{fig:ntv-vel}
\end{figure}

We extend our consideration of non-thermal velocities and radial velocities of plasma in individual active regions to the active region scale, and consider the median average and IQR spread of velocities for each active region in our study in Figure~\ref{fig:ntv-vel}. We see a similar relationship to those in those previous figures (Figures \ref{fig:vel_ntv_hist_spo}, \ref{fig:vel_ntv_hist_dec}, \ref{fig:vel_ntv_hist_dis} and \ref{fig:vel_ntv_hist_fil}), where higher non-thermal velocities are associated with upflowing plasma. We see that active regions tend to possess a relatively net-zero median average radial velocity up to around $V_{nt}\sim$15~km/s, beyond which the median average radial velocities begin a skew towards upflowing plasma. We find a moderate to strong anti-correlation between the median non-thermal velocity and median radial velocity of active regions of $r$=-0.63. We additionally considered this relationship for the leading and trailing polarities of the active regions but found no clear distinction between them.

\begin{figure}
  \centering
  \includegraphics[width=\linewidth]{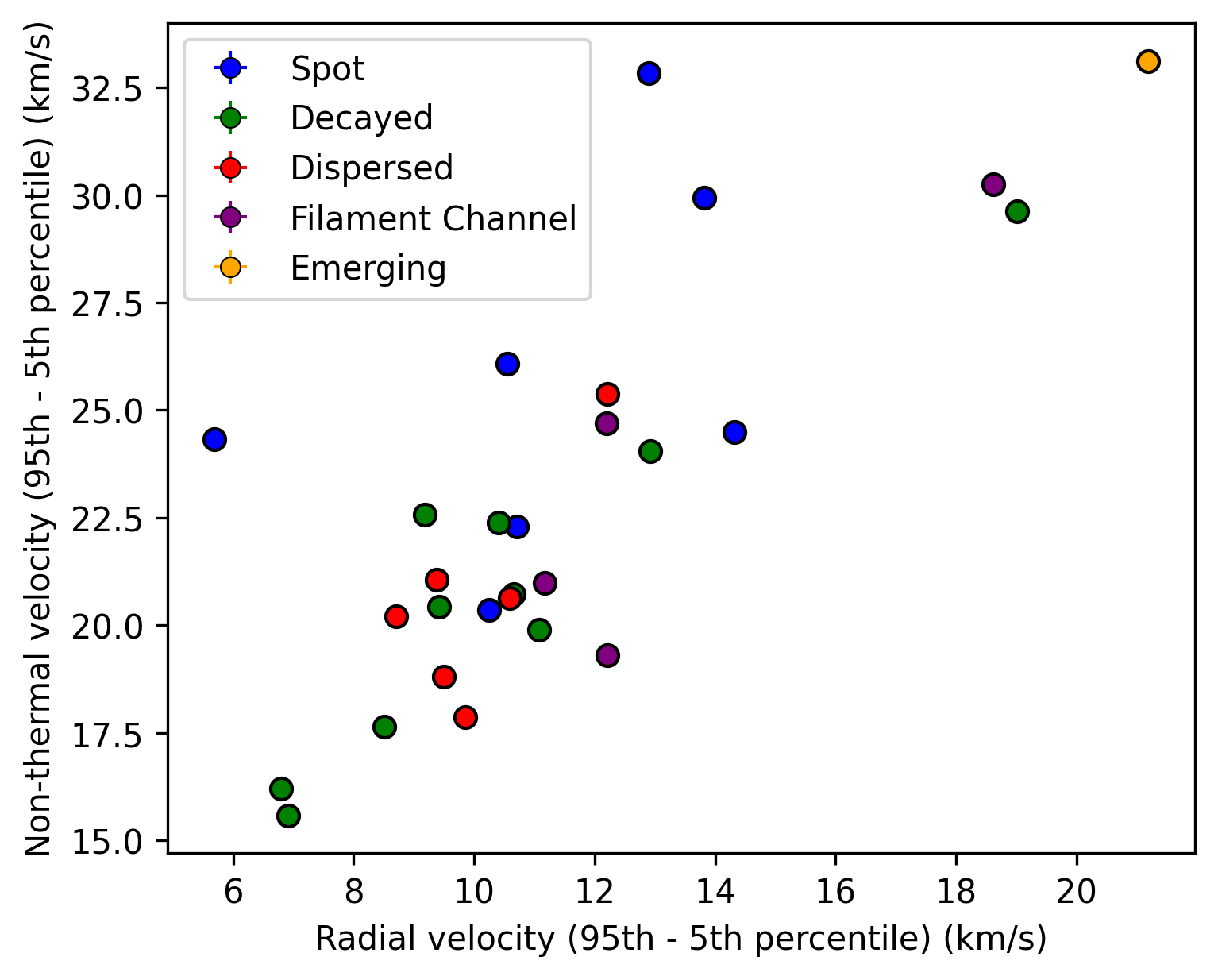}
  \caption{Difference between the 5th and 95th percentiles of radial velocity (derived from Doppler velocity) against the same range of non-thermal velocity values.\\
  Alt text: Scatter plot showing the relationship between the range (5th–95th percentile) of radial velocities and the corresponding range of non-thermal velocities for each active region. Points show a strong positive correlation between the two spreads.}
  \label{fig:vel_percentile_diff_vs_ntv_percentile_diff}
\end{figure}

As we saw in Figures \ref{fig:vel_ntv_hist_spo}, \ref{fig:vel_ntv_hist_dec}, \ref{fig:vel_ntv_hist_dis} and \ref{fig:vel_ntv_hist_fil}, at increasing non-thermal velocity values the spread of radial velocities increases largely symmetrically and then with a bias towards upflowing plasma, i.e., as non-thermal velocity values reach higher values in active regions, the spread of radial velocities increases. We plot such a trend in Figure~\ref{fig:vel_percentile_diff_vs_ntv_percentile_diff}, using the range between the 5th and 95th percentiles of the data to better capture tails in the distribution of upflowing plasma as seen in Figure~\ref{fig:ntv_vel_hists}. We see a strong positive relationship between the increase in the spread of non-thermal velocity values and the spread of radial velocities observed in an active region, with $r$=0.78.

\subsection{Non-thermal velocity versus FIP Bias}

\begin{figure}
    \centering
    \resizebox{\hsize}{!}{\includegraphics{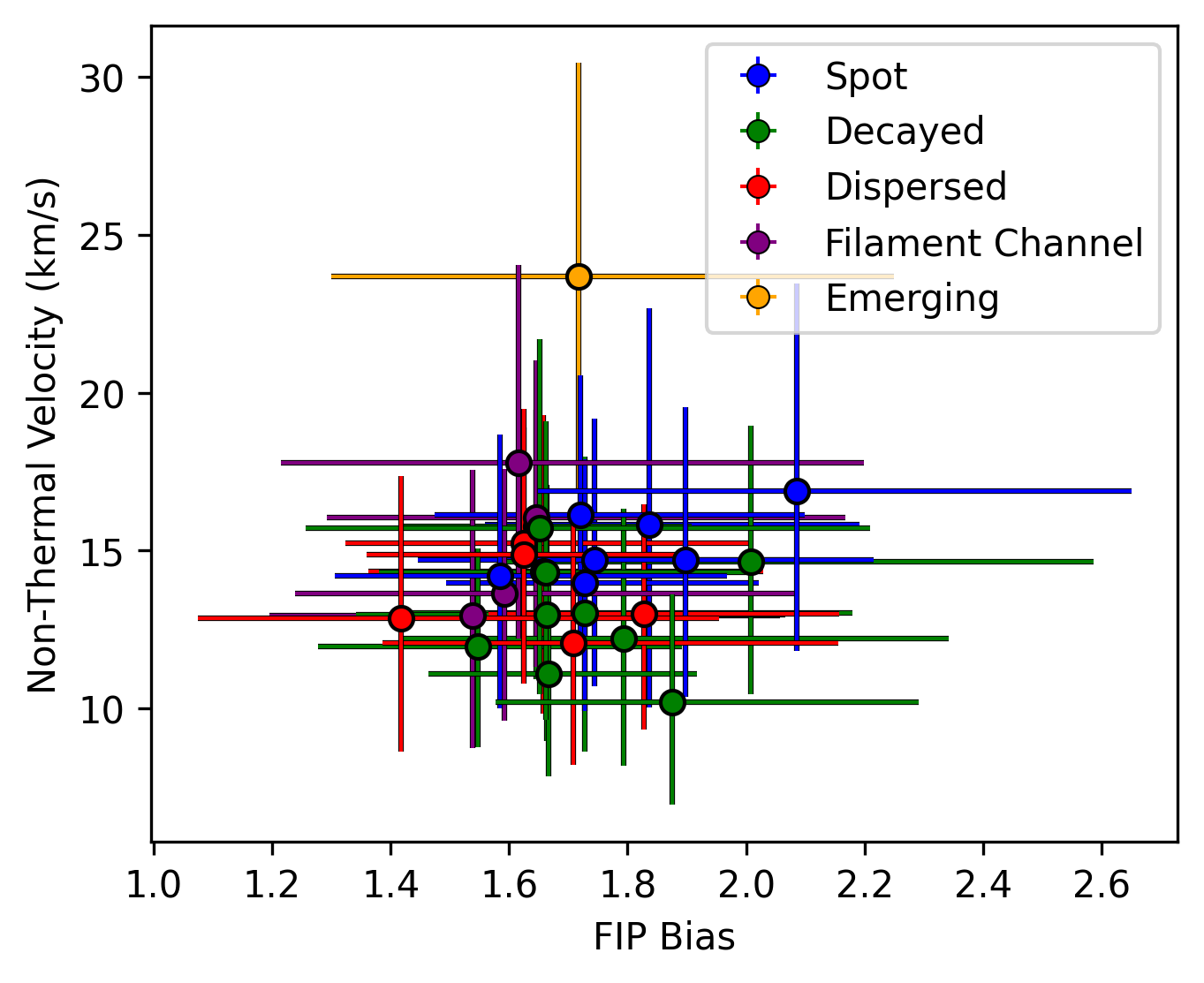}}
    \caption{The relationship between non-thermal velocity and FIP bias for all active regions. The median values are shown with circles, and the spread between the 25th and 75th percentiles is also indicated.\\
    Alt text: Scatter plot of median non-thermal velocity versus median FIP bias for all active regions, with percentile bars. The plot shows no correlation between the two.}
    \label{fig:ntv-fip}
\end{figure}

We examine the relationship between non-thermal velocity and FIP bias in Figure \ref{fig:ntv-fip}. We see no compelling correlation between the two, with moderate median non-thermal velocity values of around 15 km/s corresponding to the entire range of median FIP bias values from 1.4 to 2.1. The relationship between FIP bias and non-thermal velocity in the leading and trailing polarities shows similar results.

\section{Discussion}\label{sec:discussion}

\subsection{Magnetic flux as a driver of non-thermal motions}

As we covered in the introduction, observations of non-thermal velocity are thought to be an indicator of MHD waves and/or nanoflares (it is unclear to which degree each mechanism contributes), and previous work has shown simulations of both these mechanisms are capable of reproducing observed non-thermal velocity values in active regions for plasma at coronal temperatures (log~$T$$\sim$6.2--6.3) \citep{pontin_non-thermal_2020,asgari-targhi_observations_2024}. As we have also seen in the literature that these mechanisms are both caused by photospheric magnetic flux, we consider the relationship between magnetic flux and non-thermal velocity across our active regions to validate these relationships.

We find a moderate relationship between the non-thermal velocity of the coronal plasma (at log~$T\sim6.2$) of an active region and the underlying photospheric magnetic field strength. This indicates that regions with stronger photospheric fields host stronger coronal heating processes, or a higher filling factor of such processes.

Our findings are in agreement with those of \cite{harra_non-thermal_2012} who found a correlation between photospheric magnetic field turbulence (the random motions which are thought to cause magnetic field braiding) and log~$T\sim6.2$ non-thermal velocity, indicative of a causal energy release relationship. As work since that of \cite{harra_non-thermal_2012} has shown a quantitive relationship between photospheric magnetic flux and upward-directed Poynting flux \citep{welsch_photospheric_2015}, it is encouraging therefore that we also find a positive correlation between what is happening in the magnetic field at the photosphere and what we see non-thermally in the coronal plasma.

While \cite{brooks_measurements_2016} did not find a correlation between photospheric magnetic flux and coronal non-thermal velocity, their study considered the relationship between photospheric flux and non-thermal velocity in sections of coronal loops. \cite{harra_non-thermal_2012} considered the correlation between non-thermal velocities from across an entire active region and magnetic flux, as we do, and did find a relationship which they considered indicative of nanoflare heating. Previous studies have shown that non-thermal velocities are typically stronger around active region loop footpoints and the upflows at the active region boundaries. These features contribute to measurements of the whole active region. Indeed \cite{mondal_spatial_2025} found that different frequencies of nanoflare appear to heat plasma in different structures of an active region. Therefore, contributions from different heating mechanisms across the active region in locations other than just the loops would be expected to also affect the correlation.

Nanoflare heating is expected to initially occur at log~$T\gtrsim7$ \citep[e.g.,][]{ishikawa_detection_2017}, and it is not clear whether the signatures of such high temperature heating are still present in the log~$T\sim6.2$ plasma we consider here. The correlation we find could indicate such signatures, or could instead be related to MHD wave heating, which should occur closer to log~$T\sim6.2$ temperatures \citep{van_doorsselaere_coronal_2020,morton_alfvenic_2023}.

When we separate active region leading and trailing polarities in our analysis, we find no clear distinction between them. This is noteworthy as leading polarities are typically more coherent than comparably more dispersed trailing polarities \citep[e.g.,][]{fan_origin_1993}. It might be expected that a more fragmented trailing morphology would result in different footpoint shuffling statistics (and thus different braiding heating rates) or different wave heating rates compared to the stable leading spots. However, our results show that the relationship between non-thermal velocity and magnetic flux remains consistent regardless of polarity. This implies that the mechanism driving the excess broadening depends primarily on the local photospheric magnetic flux density, rather than the global topological coherence of the magnetic flux concentration, at least at the level of difference in coherence between leading and trailing polarities. We discuss this further, comparing spot and decayed active region types, in Section~\ref{sec:discussion_ntv_stage}.

\subsection{Connection between non-thermal broadening and bulk flows}

The trend of strongly blue-shifted plasma being associated with higher values of non-thermal velocity has been reported many times before, and reviewed extensively by e.g. \cite{doschek_dynamics_2012}. \cite{hara_coronal_2008} suggested that this might be due to unresolved upflows at high velocities, a theory largely accepted to explain this relationship \citep{del_zanna_solar_2018}. 

We find that this relationship holds for spot-type, decayed-type, dispersed-type and filament channel-type active regions. This is in agreement with \cite{yardley_widespread_2021} who, through a study of 12 active regions, proposed that such high-velocity upflows appear a common feature of all active regions. Their study relied on asymmetric line profiles, and similar analysis of such asymmetries has been performed before \citep[e.g.,][]{de_pontieu_observing_2009}. While extreme care needs to be taken to avoid introducing systematic errors from the likely asymmetric point spread function of Hinode/EIS \citep{ugarte-urra_eis_2010,warren_spectroscopic_2018}, such a physical asymmetric profile makes physical sense in the case of strong upflows.

\cite{mcintosh_alfvenic_2011} found observational evidence of Alfvénic motions apparently sufficient to heat the quiet corona and accelerate the solar wind, and work of \cite{brooks_establishing_2011} separately established a connection between active region outflows and the solar wind. As it has been shown that Alfvén waves can cause excess spectral line broadening, this ability of Alfvén waves to drive the solar wind through upflowing plasma could be the physical mechanism behind the non-thermal velocity-Doppler velocity relationship we see in Hinode/EIS data in active regions.

As pointed out by \cite{doschek_dynamics_2012}, Hinode/EIS does not quite possess the spatial or spectral resolution to resolve key processes which are responsible for these trends but that SOLAR-C seemed a promising future mission to help address our questions about line broadening. Thirteen years later, SOLAR-C/EUVST is under manufacture and is expected to launch in $\sim$2028 \citep{shimizu_solar-c_2020}. Forward modelling results show it will be capable of excellent resolution of fundamental flows and structures \citep{mckevitt_pre-flare_2026}.

\subsection{Non-thermal broadening with age and evolutionary stage}\label{sec:discussion_ntv_stage}

We had considered in our introduction whether because of the dispersal of an active region's magnetic field as it ages \citep{van_driel-gesztelyi_evolution_2015}, and because this would lead to a reduction in Poynting flux available for coronal heating, we would observe a reduction of non-thermal velocity in older active regions. However we found no relationship between non-thermal velocity and active region age. What we did see was that decayed active regions showed lower photospheric magnetic flux and lower non-thermal velocities compared to spot-type active regions (Figure~\ref{fig:ntv-mag}). Active regions are known to have different lifetimes \citep[e.g.,][]{ugarte-urra_magnetic_2015}, and so comparing their ages may not be as revealing as comparing their evolutionary stages. Indeed, the comparison in Figure~\ref{fig:ntv-mag} between spot- and decayed-type active regions supports this. This suggests a future study considering non-thermal velocity against an active region's progress along its evolutionary cycle may yield a more compelling relationship.

\subsection{Elemental fractionation with non-thermal broadening}

\cite{baker_plasma_2013} found a correlation of $r$=0.36 between the non-thermal velocity and FIP of plasma in an active region, indicating some connection between the processes causing elemental fractionation and non-thermal broadening. The pondromotive force model proposes MHD waves are responsible for this elemental fractionation \citep{laming_fip_2015}, where such MHD waves can be seen in non-thermal broadening \citep[e.g.,][]{asgari-targhi_observations_2024}. These MHD waves may be driven by nanoflares in the corona \citep{laming_first_2017}, or by photospheric motions \citep{martinez-sykora_impact_2023}.

While our Figure~\ref{fig:ntv-fip} bears some resemblance to the comparable plot for an active region of \cite{baker_plasma_2013}, we find that on active region scales there appears to be no compelling link between the non-thermal broadening and FIP bias of our active regions. We therefore suggest that the processes governing low-FIP enrichment (e.g. ponderomotive force fractionation) and those setting non-thermal widths at Fe~XII formation temperatures are not tightly coupled on active region scales. This appears consistent with studies that distinguish active region cores and upflow regions by composition and dynamics \citep[e.g.,][]{brooks_establishing_2011}.

\section{Summary}

The trends we observe across 28 active regions show a potential relationship between active region-scale coronal non-thermal velocity and photospheric magnetic flux, supporting a picture in which larger unsigned photospheric magnetic flux sustains larger Poynting flux into the corona, causing coronal heating through magnetic field braiding nanoflares and/or MHD wave heating. The moderate correlation we find is by no means conclusive, but is higher than could be expected exclusively from nanoflares given we consider active region plasma at log~$T\sim6.2$ coronal temperatures. This moderate correlation could therefore indicate active region plasma retains information about high-temperature impulsive heating, or that its heating is due more to lower-temperature MHD wave activity. The full disk observations made by Hinode/EIS enable such statistical studies and should be continued for future missions like SOLAR-C/EUVST, where a higher spatial resolution is expected \citep{mckevitt_pre-flare_2026} and will allow us to better disentangle the mechanisms behind non-thermal emission line broadening.

\begin{ack}
We thank the reviewer for their comments which greatly improved this manuscript. J.M. was supported by STFC PhD Studentship number ST/X508858/1. Data processing and analysis were performed on Austrian Scientific Computing (ASC) infrastructure (\url{https://asc.ac.at/}). S.M. and D.B. acknowledge support from UKSA grant No. UKRI920. S.M. was also supported by ESA Contract 1420 No. 4000141160/23/NL/IB, and S.M. and H.A.S.R. are supported by grant 2022-25 ST/W001004/1. S.M. and D.B. are funded under the Hinode Ops Continuation 2022-25 grant number ST/X002063/1, and D.B. and H.A.S.R. are funded under Solar Orbiter EUI Operations grant number ST/X002012/1 and UKRI980. The work of D.H.B. and I.U.U. was funded by the NASA Hinode program. This research used version 6.0.2 \citep{mumford_sunpy_2024} of the SunPy open source software package \citep{the_sunpy_community_sunpy_2020}. Data processing and analysis were performed on Austrian Scientific Computing (ASC) infrastructure.
\end{ack}

\bibliographystyle{aa}
\bibliography{references}

@article{del_zanna_solar_2018,
    title = {Solar {UV} and {X}-ray spectral diagnostics},
    volume = {15},
    issn = {2367-3648, 1614-4961},
    url = {http://link.springer.com/10.1007/s41116-018-0015-3},
    doi = {10.1007/s41116-018-0015-3},
    language = {en},
    number = {1},
    urldate = {2025-09-02},
    journal = {Living Reviews in Solar Physics},
    author = {Del Zanna, Giulio and Mason, Helen E.},
    month = dec,
    year = {2018},
    pages = {5},
}

@article{parker_nanoflares_1988,
    title = {Nanoflares and the solar {X}-ray corona},
    volume = {330},
    issn = {0004-637X, 1538-4357},
    url = {http://adsabs.harvard.edu/doi/10.1086/166485},
    doi = {10.1086/166485},
    language = {en},
    urldate = {2025-04-18},
    journal = {The Astrophysical Journal},
    author = {Parker, E. N.},
    month = jul,
    year = {1988},
    pages = {474},
}

@article{parker_magnetic_1983,
    title = {Magnetic {Neutral} {Sheets} in {Evolving} {Fields} - {Part} {Two} - {Formation} of the {Solar} {Corona}},
    volume = {264},
    issn = {0004-637X, 1538-4357},
    url = {http://adsabs.harvard.edu/doi/10.1086/160637},
    doi = {10.1086/160637},
    language = {en},
    urldate = {2025-08-29},
    journal = {The Astrophysical Journal},
    author = {Parker, E. N.},
    month = jan,
    year = {1983},
    pages = {642},
}

@article{yeates_coronal_2014,
    title = {The coronal energy input from magnetic braiding},
    volume = {564},
    issn = {0004-6361, 1432-0746},
    url = {http://www.aanda.org/10.1051/0004-6361/201323276},
    doi = {10.1051/0004-6361/201323276},
    urldate = {2025-08-25},
    journal = {Astronomy \& Astrophysics},
    author = {Yeates, A. R. and Bianchi, F. and Welsch, B. T. and Bushby, P. J.},
    month = apr,
    year = {2014},
    pages = {A131},
}

@article{viall_evidence_2012,
    title = {Evidence for widespread cooling in an active region observed with the sdo atmospheric imaging assembly},
    volume = {753},
    issn = {0004-637X, 1538-4357},
    url = {https://iopscience.iop.org/article/10.1088/0004-637X/753/1/35},
    doi = {10.1088/0004-637X/753/1/35},
    number = {1},
    urldate = {2025-08-29},
    journal = {The Astrophysical Journal},
    author = {Viall, Nicholeen M. and Klimchuk, James A.},
    month = jul,
    year = {2012},
    pages = {35},
}

@article{ishikawa_detection_2017,
    title = {Detection of nanoflare-heated plasma in the solar corona by the {FOXSI}-2 sounding rocket},
    volume = {1},
    issn = {2397-3366},
    url = {https://www.nature.com/articles/s41550-017-0269-z},
    doi = {10.1038/s41550-017-0269-z},
    language = {en},
    number = {11},
    urldate = {2025-08-29},
    journal = {Nature Astronomy},
    author = {Ishikawa, Shin-nosuke and Glesener, Lindsay and Krucker, Säm and Christe, Steven and Buitrago-Casas, Juan Camilo and Narukage, Noriyuki and Vievering, Juliana},
    month = oct,
    year = {2017},
    pages = {771--774},
}

@article{klimchuk_key_2015,
    title = {Key aspects of coronal heating},
    volume = {373},
    issn = {1364-503X, 1471-2962},
    url = {https://royalsocietypublishing.org/doi/10.1098/rsta.2014.0256},
    doi = {10.1098/rsta.2014.0256},
    language = {en},
    number = {2042},
    urldate = {2025-08-29},
    journal = {Philosophical Transactions of the Royal Society A: Mathematical, Physical and Engineering Sciences},
    author = {Klimchuk, James A.},
    month = may,
    year = {2015},
    pages = {20140256},
}

@article{aschwanden_coronal_2007,
    title = {The {Coronal} {Heating} {Paradox}},
    volume = {659},
    issn = {0004-637X, 1538-4357},
    url = {https://iopscience.iop.org/article/10.1086/513070},
    doi = {10.1086/513070},
    language = {en},
    number = {2},
    urldate = {2025-08-29},
    journal = {The Astrophysical Journal},
    author = {Aschwanden, Markus J. and Winebarger, Amy and Tsiklauri, David and Peter, Hardi},
    month = apr,
    year = {2007},
    pages = {1673--1681},
}

@article{de_pontieu_observing_2009,
    title = {Observing the roots of solar coronal heating—in the chromosphere},
    volume = {701},
    issn = {0004-637X, 1538-4357},
    url = {https://iopscience.iop.org/article/10.1088/0004-637X/701/1/L1},
    doi = {10.1088/0004-637X/701/1/L1},
    number = {1},
    urldate = {2025-08-29},
    journal = {The Astrophysical Journal},
    author = {De Pontieu, Bart and McIntosh, Scott W. and Hansteen, Viggo H. and Schrijver, Carolus J.},
    month = aug,
    year = {2009},
    pages = {L1--L6},
}

@article{klimchuk_are_2014,
    title = {Are chromospheric nanoflares a primary source of coronal plasma?},
    volume = {791},
    copyright = {http://iopscience.iop.org/info/page/text-and-data-mining},
    issn = {0004-637X, 1538-4357},
    url = {https://iopscience.iop.org/article/10.1088/0004-637X/791/1/60},
    doi = {10.1088/0004-637X/791/1/60},
    number = {1},
    urldate = {2025-08-29},
    journal = {The Astrophysical Journal},
    author = {Klimchuk, J. A. and Bradshaw, S. J.},
    month = jul,
    year = {2014},
    pages = {60},
}

@article{patsourakos_core_2014,
    title = {Core and wing densities of asymmetric coronal spectral profiles: implications for the mass supply of the solar corona},
    volume = {781},
    copyright = {http://iopscience.iop.org/info/page/text-and-data-mining},
    issn = {0004-637X, 1538-4357},
    shorttitle = {Core and wing densities of asymmetric coronal spectral profiles},
    url = {https://iopscience.iop.org/article/10.1088/0004-637X/781/2/58},
    doi = {10.1088/0004-637X/781/2/58},
    number = {2},
    urldate = {2025-08-29},
    journal = {The Astrophysical Journal},
    author = {Patsourakos, S. and Klimchuk, J. A. and Young, P. R.},
    month = jan,
    year = {2014},
    pages = {58},
}

@article{mcintosh_alfvenic_2011,
    title = {Alfvénic waves with sufficient energy to power the quiet solar corona and fast solar wind},
    volume = {475},
    copyright = {http://www.springer.com/tdm},
    issn = {0028-0836, 1476-4687},
    url = {https://www.nature.com/articles/nature10235},
    doi = {10.1038/nature10235},
    language = {en},
    number = {7357},
    urldate = {2025-08-25},
    journal = {Nature},
    author = {McIntosh, Scott W. and De Pontieu, Bart and Carlsson, Mats and Hansteen, Viggo and Boerner, Paul and Goossens, Marcel},
    month = jul,
    year = {2011},
    pages = {477--480},
}

@article{van_doorsselaere_coronal_2020,
    title = {Coronal {Heating} by {MHD} {Waves}},
    volume = {216},
    issn = {0038-6308, 1572-9672},
    url = {http://link.springer.com/10.1007/s11214-020-00770-y},
    doi = {10.1007/s11214-020-00770-y},
    language = {en},
    number = {8},
    urldate = {2025-08-29},
    journal = {Space Science Reviews},
    author = {Van Doorsselaere, Tom and Srivastava, Abhishek K. and Antolin, Patrick and Magyar, Norbert and Vasheghani Farahani, Soheil and Tian, Hui and Kolotkov, Dmitrii and Ofman, Leon and Guo, Mingzhe and Arregui, Iñigo and De Moortel, Ineke and Pascoe, David},
    month = dec,
    year = {2020},
    pages = {140},
}

@article{welsch_photospheric_2015,
    title = {The photospheric {Poynting} flux and coronal heating},
    volume = {67},
    issn = {2053-051X, 0004-6264},
    url = {https://academic.oup.com/pasj/article/doi/10.1093/pasj/psu151/1512951},
    doi = {10.1093/pasj/psu151},
    language = {en},
    number = {2},
    urldate = {2025-08-24},
    journal = {Publications of the Astronomical Society of Japan},
    author = {Welsch, Brian T.},
    month = apr,
    year = {2015},
    pages = {18},
}

@article{culhane_euv_2007,
    title = {The {EUV} {Imaging} {Spectrometer} for {Hinode}},
    volume = {243},
    copyright = {http://www.springer.com/tdm},
    issn = {0038-0938, 1573-093X},
    url = {http://link.springer.com/10.1007/s01007-007-0293-1},
    doi = {10.1007/s01007-007-0293-1},
    language = {en},
    number = {1},
    urldate = {2025-01-06},
    journal = {Solar Physics},
    author = {Culhane, J. L. and Harra, L. K. and James, A. M. and Al-Janabi, K. and Bradley, L. J. and Chaudry, R. A. and Rees, K. and Tandy, J. A. and Thomas, P. and Whillock, M. C. R. and Winter, B. and Doschek, G. A. and Korendyke, C. M. and Brown, C. M. and Myers, S. and Mariska, J. and Seely, J. and Lang, J. and Kent, B. J. and Shaughnessy, B. M. and Young, P. R. and Simnett, G. M. and Castelli, C. M. and Mahmoud, S. and Mapson-Menard, H. and Probyn, B. J. and Thomas, R. J. and Davila, J. and Dere, K. and Windt, D. and Shea, J. and Hagood, R. and Moye, R. and Hara, H. and Watanabe, T. and Matsuzaki, K. and Kosugi, T. and Hansteen, V. and Wikstol, Ø.},
    month = sep,
    year = {2007},
    pages = {19--61},
}

@article{kosugi_hinode_2007,
    title = {The {Hinode} ({Solar}-{B}) {Mission}: {An} {Overview}},
    volume = {243},
    copyright = {http://www.springer.com/tdm},
    issn = {0038-0938, 1573-093X},
    shorttitle = {The {Hinode} ({Solar}-{B}) {Mission}},
    url = {http://link.springer.com/10.1007/s11207-007-9014-6},
    doi = {10.1007/s11207-007-9014-6},
    language = {en},
    number = {1},
    urldate = {2025-01-06},
    journal = {Solar Physics},
    author = {Kosugi, T. and Matsuzaki, K. and Sakao, T. and Shimizu, T. and Sone, Y. and Tachikawa, S. and Hashimoto, T. and Minesugi, K. and Ohnishi, A. and Yamada, T. and Tsuneta, S. and Hara, H. and Ichimoto, K. and Suematsu, Y. and Shimojo, M. and Watanabe, T. and Shimada, S. and Davis, J. M. and Hill, L. D. and Owens, J. K. and Title, A. M. and Culhane, J. L. and Harra, L. K. and Doschek, G. A. and Golub, L.},
    month = sep,
    year = {2007},
    pages = {3--17},
}

@article{lorincik_plasma_2020,
    title = {Plasma {Diagnostics} from {Active} {Region} and {Quiet}-{Sun} {Spectra} {Observed} by {Hinode}/{EIS}: {Quantifying} the {Departures} from a {Maxwellian} {Distribution}},
    volume = {893},
    issn = {0004-637X, 1538-4357},
    shorttitle = {Plasma {Diagnostics} from {Active} {Region} and {Quiet}-{Sun} {Spectra} {Observed} by {Hinode}/{EIS}},
    url = {https://iopscience.iop.org/article/10.3847/1538-4357/ab8010},
    doi = {10.3847/1538-4357/ab8010},
    number = {1},
    urldate = {2025-09-02},
    journal = {The Astrophysical Journal},
    author = {Lörinčík, Juraj and Dudík, Jaroslav and Del Zanna, Giulio and Dzifčáková, Elena and Mason, Helen E.},
    month = apr,
    year = {2020},
    pages = {34},
}

@article{hassler_line_1990,
    title = {Line {Broadening} of {MG} {X} lambda lambda 609 and 625 {Coronal} {Emission} {Lines} {Observed} above the {Solar} {Limb}},
    volume = {348},
    issn = {0004-637X},
    url = {https://ui.adsabs.harvard.edu/abs/1990ApJ...348L..77H},
    doi = {10.1086/185635},
    urldate = {2025-09-02},
    journal = {The Astrophysical Journal},
    author = {Hassler, Donald M. and Rottman, Gary J. and Shoub, Edward C. and Holzer, Thomas E.},
    month = jan,
    year = {1990},
    note = {ADS Bibcode: 1990ApJ...348L..77H},
    pages = {L77},
}

@article{hara_coronal_2008,
    title = {Coronal {Plasma} {Motions} near {Footpoints} of {Active} {Region} {Loops} {Revealed} from {Spectroscopic} {Observations} with \textit{{Hinode}} {EIS}},
    volume = {678},
    issn = {0004-637X, 1538-4357},
    url = {https://iopscience.iop.org/article/10.1086/588252},
    doi = {10.1086/588252},
    language = {en},
    number = {1},
    urldate = {2025-09-02},
    journal = {The Astrophysical Journal},
    author = {Hara, Hirohisa and Watanabe, Tetsuya and Harra, Louise K. and Culhane, J. Leonard and Young, Peter R. and Mariska, John T. and Doschek, George A.},
    month = may,
    year = {2008},
    pages = {L67--L71},
}

@article{bradshaw_self-consistent_2003,
    title = {A self-consistent treatment of radiation in coronal loop modelling},
    volume = {401},
    issn = {0004-6361, 1432-0746},
    url = {http://www.aanda.org/10.1051/0004-6361:20030089},
    doi = {10.1051/0004-6361:20030089},
    number = {2},
    urldate = {2025-09-02},
    journal = {Astronomy \& Astrophysics},
    author = {Bradshaw, S. J. and Mason, H. E.},
    month = apr,
    year = {2003},
    pages = {699--709},
}

@article{dzifcakova_diagnostics_2011,
    title = {Diagnostics of the \textit{κ} -distribution using {Si} {III} lines in the solar transition region},
    volume = {531},
    issn = {0004-6361, 1432-0746},
    url = {http://www.aanda.org/10.1051/0004-6361/201016287},
    doi = {10.1051/0004-6361/201016287},
    urldate = {2025-09-02},
    journal = {Astronomy \& Astrophysics},
    author = {Dzifčáková, E. and Kulinová, A.},
    month = jul,
    year = {2011},
    pages = {A122},
}

@article{patsourakos_nonthermal_2006,
    title = {Nonthermal {Spectral} {Line} {Broadening} and the {Nanoflare} {Model}},
    volume = {647},
    issn = {0004-637X, 1538-4357},
    url = {https://iopscience.iop.org/article/10.1086/505517},
    doi = {10.1086/505517},
    language = {en},
    number = {2},
    urldate = {2025-09-02},
    journal = {The Astrophysical Journal},
    author = {Patsourakos, S. and Klimchuk, J. A.},
    month = aug,
    year = {2006},
    pages = {1452--1465},
}

@article{banerjee_signatures_2009,
    title = {Signatures of {Alfvén} waves in the polar coronal holes as seen by {EIS}/{Hinode}},
    volume = {501},
    issn = {0004-6361, 1432-0746},
    url = {http://www.aanda.org/10.1051/0004-6361/200912242},
    doi = {10.1051/0004-6361/200912242},
    number = {3},
    urldate = {2025-09-02},
    journal = {Astronomy \& Astrophysics},
    author = {Banerjee, D. and Pérez-Suárez, D. and Doyle, J. G.},
    month = jul,
    year = {2009},
    pages = {L15--L18},
}

@article{bradshaw_reconnection-driven_2011,
    title = {A {Reconnection}-{Driven} {Rarefaction} {Wave} {Model} for {Coronal} {Outflows}},
    volume = {743},
    issn = {0004-637X, 1538-4357},
    url = {https://iopscience.iop.org/article/10.1088/0004-637X/743/1/66},
    doi = {10.1088/0004-637X/743/1/66},
    number = {1},
    urldate = {2025-09-02},
    journal = {The Astrophysical Journal},
    author = {Bradshaw, S. J. and Aulanier, G. and Del Zanna, G.},
    month = dec,
    year = {2011},
    pages = {66},
}

@article{feldman_unresolved_1983,
    title = {On the unresolved fine structures of the solar atmosphere in the 30,000-200,000 {K} temperature region},
    volume = {275},
    issn = {0004-637X},
    url = {https://ui.adsabs.harvard.edu/abs/1983ApJ...275..367F},
    doi = {10.1086/161539},
    urldate = {2025-09-02},
    journal = {The Astrophysical Journal},
    author = {Feldman, U.},
    month = dec,
    year = {1983},
    note = {ADS Bibcode: 1983ApJ...275..367F},
    pages = {367--373},
}

@article{brooks_measurements_2016,
    title = {Measurements of {Non}-{Thermal} {Line} {Widths} in {Solar} {Active} {Regions}},
    volume = {820},
    issn = {0004-637X, 1538-4357},
    url = {https://iopscience.iop.org/article/10.3847/0004-637X/820/1/63},
    doi = {10.3847/0004-637X/820/1/63},
    number = {1},
    urldate = {2025-08-29},
    journal = {The Astrophysical Journal},
    author = {Brooks, David H. and Warren, Harry P.},
    month = mar,
    year = {2016},
    pages = {63},
}

@article{pontin_non-thermal_2020,
    title = {Non-thermal line broadening due to braiding-induced turbulence in solar coronal loops},
    volume = {639},
    copyright = {https://www.edpsciences.org/en/authors/copyright-and-licensing},
    issn = {0004-6361, 1432-0746},
    url = {https://www.aanda.org/10.1051/0004-6361/202037582},
    doi = {10.1051/0004-6361/202037582},
    urldate = {2025-08-24},
    journal = {Astronomy \& Astrophysics},
    author = {Pontin, D. I. and Peter, H. and Chitta, L. P.},
    month = jul,
    year = {2020},
    pages = {A21},
}

@article{asgari-targhi_observations_2024,
    title = {Observations of {Nonthermal} {Velocities} and {Comparisons} with an {Alfvén} {Wave} {Turbulence} {Model} in {Solar} {Active} {Regions}},
    volume = {968},
    issn = {0004-637X, 1538-4357},
    url = {https://iopscience.iop.org/article/10.3847/1538-4357/ad434a},
    doi = {10.3847/1538-4357/ad434a},
    number = {1},
    urldate = {2025-04-27},
    journal = {The Astrophysical Journal},
    author = {Asgari-Targhi, M. and Brooks, D. H. and Hahn, M. and Imada, S. and Tajfirouze, E. and Savin, D. W.},
    month = jun,
    year = {2024},
    pages = {7},
}

@article{laming_fip_2015,
    title = {The {FIP} and {Inverse} {FIP} {Effects} in {Solar} and {Stellar} {Coronae}},
    volume = {12},
    issn = {2367-3648, 1614-4961},
    url = {http://link.springer.com/10.1007/lrsp-2015-2},
    doi = {10.1007/lrsp-2015-2},
    language = {en},
    number = {1},
    urldate = {2025-08-26},
    journal = {Living Reviews in Solar Physics},
    author = {Laming, J. Martin},
    month = dec,
    year = {2015},
    pages = {2},
}

@article{laming_non-wkb_2012,
    title = {Non-{WKB} {Models} of the {First} {Ionization} {Potential} {Effect}: {The} {Role} of {Slow} {Mode} {Waves}},
    volume = {744},
    issn = {0004-637X, 1538-4357},
    shorttitle = {Non-wkb models of the first ionization potential effect},
    url = {https://iopscience.iop.org/article/10.1088/0004-637X/744/2/115},
    doi = {10.1088/0004-637X/744/2/115},
    number = {2},
    urldate = {2025-09-02},
    journal = {The Astrophysical Journal},
    author = {Laming, J. Martin},
    month = jan,
    year = {2012},
    pages = {115},
}

@article{baker_plasma_2013,
    title = {Plasma {Composition} in a {Sigmoidal} {Anemone} {Active} {Region}},
    volume = {778},
    copyright = {http://iopscience.iop.org/info/page/text-and-data-mining},
    issn = {0004-637X, 1538-4357},
    url = {https://iopscience.iop.org/article/10.1088/0004-637X/778/1/69},
    doi = {10.1088/0004-637X/778/1/69},
    number = {1},
    urldate = {2025-08-25},
    journal = {The Astrophysical Journal},
    author = {Baker, D. and Brooks, D. H. and Démoulin, P. and Van Driel-Gesztelyi, L. and Green, L. M. and Steed, K. and Carlyle, J.},
    month = nov,
    year = {2013},
    pages = {69},
}

@article{mihailescu_what_2022,
    title = {What {Determines} {Active} {Region} {Coronal} {Plasma} {Composition}?},
    volume = {933},
    issn = {0004-637X, 1538-4357},
    url = {https://iopscience.iop.org/article/10.3847/1538-4357/ac6e40},
    doi = {10.3847/1538-4357/ac6e40},
    language = {en},
    number = {2},
    urldate = {2024-12-11},
    journal = {The Astrophysical Journal},
    author = {Mihailescu, Teodora and Baker, Deborah and Green, Lucie M. and Van Driel-Gesztelyi, Lidia and Long, David M. and Brooks, David H. and To, Andy S. H.},
    month = jul,
    year = {2022},
    pages = {245},
}

@article{brooks_full-sun_2015,
    title = {Full-{Sun} observations for identifying the source of the slow solar wind},
    volume = {6},
    issn = {2041-1723},
    url = {https://www.nature.com/articles/ncomms6947},
    doi = {10.1038/ncomms6947},
    language = {en},
    number = {1},
    urldate = {2025-01-06},
    journal = {Nature Communications},
    author = {Brooks, David H. and Ugarte-Urra, Ignacio and Warren, Harry P.},
    month = jan,
    year = {2015},
    pages = {5947},
}

@article{harra_non-thermal_2012,
    title = {Non-{Thermal} {Response} of the {Corona} to the {Magnetic} {Flux} {Dispersal} in the {Photosphere} of a {Decaying} {Active} {Region}},
    volume = {759},
    issn = {0004-637X, 1538-4357},
    url = {https://iopscience.iop.org/article/10.1088/0004-637X/759/2/104},
    doi = {10.1088/0004-637X/759/2/104},
    number = {2},
    urldate = {2025-08-25},
    journal = {The Astrophysical Journal},
    author = {Harra, L. K. and Abramenko, V. I.},
    month = nov,
    year = {2012},
    pages = {104},
}

@article{mondal_spatial_2025,
    title = {Spatial and {Temporal} {Distribution} of {Nanoflare} {Heating} during {Active} {Region} {Evolution}},
    volume = {980},
    issn = {0004-637X, 1538-4357},
    url = {https://iopscience.iop.org/article/10.3847/1538-4357/ada3d6},
    doi = {10.3847/1538-4357/ada3d6},
    number = {1},
    urldate = {2025-09-03},
    journal = {The Astrophysical Journal},
    author = {Mondal, Biswajit and Klimchuk, James A. and Winebarger, Amy R. and Athiray, P. S. and Liu, Jiayi},
    month = feb,
    year = {2025},
    pages = {75},
}

@article{doschek_dynamics_2012,
    title = {The {Dynamics} and {Heating} of {Active} {Region} {Loops}},
    volume = {754},
    issn = {0004-637X},
    url = {https://ui.adsabs.harvard.edu/abs/2012ApJ...754..153D},
    doi = {10.1088/0004-637X/754/2/153},
    urldate = {2025-09-04},
    journal = {The Astrophysical Journal},
    author = {Doschek, G. A.},
    month = aug,
    year = {2012},
    note = {ADS Bibcode: 2012ApJ...754..153D},
    pages = {153},
}

@article{yardley_widespread_2021,
    title = {Widespread occurrence of high-velocity upflows in solar active regions},
    volume = {650},
    copyright = {https://www.edpsciences.org/en/authors/copyright-and-licensing},
    issn = {0004-6361, 1432-0746},
    url = {https://www.aanda.org/10.1051/0004-6361/202141131},
    doi = {10.1051/0004-6361/202141131},
    urldate = {2025-08-25},
    journal = {Astronomy \& Astrophysics},
    author = {Yardley, S. L. and Brooks, D. H. and Baker, D.},
    month = jun,
    year = {2021},
    pages = {L10},
}

@techreport{ugarte-urra_eis_2010,
    address = {Washington, DC},
    type = {Technical {Report}},
    title = {{EIS} {Point} {Spread} {Function}},
    url = {https://solarb.mssl.ucl.ac.uk/SolarB/eis_docs/eis_notes/08_COMA/eis_swnote_08.pdf},
    language = {en},
    number = {8 (1.0)},
    institution = {Naval Research Laboratory},
    author = {Ugarte-Urra, Ignacio},
    month = dec,
    year = {2010},
}

@article{warren_spectroscopic_2018,
    title = {Spectroscopic {Observations} of {Current} {Sheet} {Formation} and {Evolution}},
    volume = {854},
    issn = {0004-637X, 1538-4357},
    url = {https://iopscience.iop.org/article/10.3847/1538-4357/aaa9b8},
    doi = {10.3847/1538-4357/aaa9b8},
    number = {2},
    urldate = {2025-04-28},
    journal = {The Astrophysical Journal},
    author = {Warren, Harry P. and Brooks, David H. and Ugarte-Urra, Ignacio and Reep, Jeffrey W. and Crump, Nicholas A. and Doschek, George A.},
    month = feb,
    year = {2018},
    pages = {122},
}

@inproceedings{shimizu_solar-c_2020,
    address = {Online Only, United States},
    title = {The {Solar}-{C} ({EUVST}) mission: the latest status},
    isbn = {978-1-5106-3675-0 978-1-5106-3676-7},
    shorttitle = {The {Solar}-{C}\_EUVST mission},
    url = {https://www.spiedigitallibrary.org/conference-proceedings-of-spie/11444/2560887/The-Solar-C_EUVST-mission-the-latest-status/10.1117/12.2560887.full},
    doi = {10.1117/12.2560887},
    urldate = {2025-05-21},
    booktitle = {Space {Telescopes} and {Instrumentation} 2020: {Ultraviolet} to {Gamma} {Ray}},
    publisher = {SPIE},
    author = {Shimizu, Toshifumi and Imada, Shinsuke and Kawate, Tomoko and Suematsu, Yoshinori and Hara, Hirohisa and Tsuzuki, Toshihiro and Katsukawa, Yukio and Kubo, Masahito and Ishikawa, Ryoko and Watanabe, Tetsuya and Toriumi, Shin and Ichimoto, Kiyoshi and Nagata, Shin'ichi and Hasegawa, Takahiro and Yokoyama, Takaaki and Watanabe, Kyoko and Tsuno, Katsuhiko and Korendyke, Clarence M. and Warren, Harry P. and De Pontieu, Bart and Boerner, Paul and Solanki, Sami K. and Teriaca, Luca and Schühle, Udo and Matthews, Sarah and Long, David and Thomas, William and Hancock, Barry and Reid, Hamish and Fludra, Andrzej and Auchere, Frederic and Andretta, Vincenzo and Naletto, Giampiero and Poletto, Luca and Harra, Louise},
    editor = {Den Herder, Jan-Willem A. and Nakazawa, Kazuhiro and Nikzad, Shouleh},
    month = dec,
    year = {2020},
    pages = {19},
}

@article{brooks_establishing_2011,
    title = {Establishing a {Connection} {Between} {Active} {Region} {Outflows} and the {Solar} {Wind}: {Abundance} {Measurements} with {EIS}/{Hinode}},
    volume = {727},
    issn = {2041-8205, 2041-8213},
    shorttitle = {Establishing a {Connection} {Between} {Active} {Region} {Outflows} and the {Solar} {Wind}},
    url = {https://iopscience.iop.org/article/10.1088/2041-8205/727/1/L13},
    doi = {10.1088/2041-8205/727/1/L13},
    language = {en},
    number = {1},
    urldate = {2025-01-06},
    journal = {The Astrophysical Journal},
    author = {Brooks, David H. and Warren, Harry P.},
    month = jan,
    year = {2011},
    pages = {L13},
}

@article{van_driel-gesztelyi_evolution_2015,
    title = {Evolution of {Active} {Regions}},
    volume = {12},
    issn = {2367-3648, 1614-4961},
    url = {http://link.springer.com/10.1007/lrsp-2015-1},
    doi = {10.1007/lrsp-2015-1},
    language = {en},
    number = {1},
    urldate = {2025-01-06},
    journal = {Living Reviews in Solar Physics},
    author = {Van Driel-Gesztelyi, Lidia and Green, Lucie May},
    month = dec,
    year = {2015},
    pages = {1},
}

@misc{mumford_sunpy_2024,
    title = {sunpy: {A} {Core} {Package} for {Solar} {Physics}},
    url = {https://doi.org/10.5281/zenodo.14292170},
    publisher = {Zenodo},
    author = {Mumford, Stuart J. and Freij, Nabil and Stansby, David and Christe, Steven and Shih, Albert Y. and Ireland, Jack and Mayer, Florian and Hughitt, V. Keith and Ryan, Daniel F. and Liedtke, Simon and Barnes, Will and Hayes, Laura and Pérez-Suárez, David and I, Vishnunarayan K. and Chakraborty, Pritish and Inglis, Andrew and Pattnaik, Punyaslok and Sipőcz, Brigitta and MacBride, Conor and Sharma, Rishabh and Leonard, Andrew and Hewett, Russell and Hamilton, Alex and Manhas, Abhijeet and Panda, Asish and Earnshaw, Matt and Choudhary, Nitin and Kumar, Ankit and Singh, Raahul and Chanda, Prateek and Haque, Md Akramul and Wilson, Alasdair and Kirk, Michael S. and Maloney, Shane and Mueller, Michael and Konge, Sudarshan and Wentzel-Long, Matt and Srivastava, Rajul and Bennett, Samuel and Jain, Yash and Zivadinovic, Lazar and Baruah, Ankit and Arbolante, Quinn and Simon, Trestan F. and Charlton, Michael and Mishra, Sashank and Paul, Jeffrey Aaron and Verma, Akash and Chorley, Nicky and Chouhan, Aryan and Gilly, Chris R. and {Himanshu} and Mason, James Paul and Modi, Sanskar and Sharma, Yash and {Naman9639} and Bobra, Monica and Tyagi, Akshit and Rozo, Jose Ivan Campos and Manley, Larry and Ivashkiv, Kateryna and Laitinen, Timo and Chatterjee, Agneet and Dixit, Ansh and Gieseler, Jan and Dulange, Jayraj and Forstner, Johan Freiherr von and Bazán, Juanjo and Stern, Kris Akira and Shukla, Aryan and Evans, John and Jain, Sarthak and Malocha, Michael and Ghosh, Sourav and {Airmansmith97} and Stańczak, Dominik and Singh, Manit and Singh, Rajiv Ranjan and Visscher, Ruben De and Verma, Shresth and Lemos, Sophie and Agrawal, Ankit and Alam, Arib and Sinha, Aritra and Graham, Brett J. and Buddhika, Dumindu and Collier, Hannah and Pathak, Himanshu and Rideout, Jai Ram and Sharma, Swapnil and Briseno, Daniel Garcia and Shah, Harsh and Park, Jongyeob and Bates, Matt and Shukla, Devansh and Giger, Marius and Mishra, Pankaj and Sharma, Deepankar and Goel, Dhruv and Taylor, Garrison and Cetusic, Goran and Reiter, Guntbert and {Jacob} and Inchaurrandieta, Mateo and Sharma, Piyush and Dacie, Sally and Dubey, Sanjeev and Eigenbrot, Arthur and Mampaey, Benjamin and Bray, Erik and Murphy, Nick and Surve, Rutuja and Zahniy, Serge and Sidhu, Sudeep and Meszaros, Tomas and Parkhi, Utkarsh and Russell, William and Bose, Abhigyan and Pandey, Abhishek and Price-Whelan, Adrian and Hossam, Ahmed and Jahagirdar, Amogh and Chicrala, André and Mishra, Aniket and {Ankit} and Guennou, Chloé and D'Avella, Daniel and Williams, Daniel and Verma, Dipanshu and Ballew, Jordan and Agrawal, Krish and Manasia, Mubin and Kulkarni, Neeraj and Singh, Nischal and Lodha, Priyank and Kooten, Samuel J. Van and Mishra, Shivansh and Robitaille, Thomas and Augspurger, Tom and Krishan, Yash and Bahuleyan, Abijith and Bhope, Adwait and Gaba, Amarjit Singh and Hill, Andrew and Wiedemann, Bernhard M. and Molina, Carlos and Keşkek, Duygu and Habib, Ishtyaq and Letts, Joseph and Singaravelan, Karthikeyan and Ranjan, Kritika and Pandey, Mridul and Altunian, Noah and Streicher, Ole and Gomillion, Reid and Agarwal, Samriddhi and Kothari, Yash and Malik, Yash and Nomiya, Yukie and Burnett, Zach and Stevens, Abigail L. and Shauryam, Akhoury and Kaszynski, Alex and Wang, Alex and Mehrotra, Ambar and Tang, Andy and Sinha, Anubhav and Smith, Arfon and Kustov, Arseniy and Bastola, Ashish and Stone, Brandon and Bard, Chris and Robert, Clément and Behn, Ed and Mansky, Ed and Arias, Emmanuel and Paganin, Enrico and Tollerud, Erik and Dover, Fionnlagh Mackenzie and Verstringe, Freek and Kdimati, Ghaith and Kumar, Gulshan and Mathur, Harsh and Babuschkin, Igor and Calixto, James and Wimbish, Jaylen and Qing, Jia and Buitrago-Casas, Juan Camilo and Krishna, Kalpesh and Chaudhari, Kaustubh and Hiware, Kaustubh and Ghosh, Koustav and McKee, Kurt and Mangaonkar, Manas and Cheung, Mark and Mendero, Matthew and Dedhia, Megh and Schoentgen, Mickaël and {Mika} and Lyes, Mouloudi Mohamed and Shahdadpuri, Nakul and Srinivasan, Naveen and Gyenge, Norbert G. and {OussCHE} and Wright, Paul J. and Mekala, Rajasekhar Reddy and Das, Ratul and Chalana, Rehan and Mishra, Rishabh and Sharma, Rohan and Badman, Samuel T. and Srikanth, Shashank and Jain, Shubham and Yu, Sijie and Hansda, Sirjan and Farah, Suleiman and Kannojia, Swapnil and Chistie, Syed Md Mihan and Qing, Tan Jia and Yadav, Tannmay and Paul, Tathagata and Wilkinson, Tessa D. and Caswell, Thomas A. and Braccia, Thomas and Pereira, Tiago M. D. and Gates, Tim and Dang, Trung Kien and Bankar, Varun and Jamieson, William and Agrawal, Yudhik and {graham} and {pradeep} and {resakra} and {yasintoda} and Attie, Raphael and Murray, Sophie A.},
    month = dec,
    year = {2024},
    doi = {10.5281/zenodo.14292170},
}

@article{the_sunpy_community_sunpy_2020,
    title = {The {SunPy} {Project}: {Open} {Source} {Development} and {Status} of the {Version} 1.0 {Core} {Package}},
    volume = {890},
    issn = {0004-637X, 1538-4357},
    shorttitle = {The {SunPy} {Project}},
    url = {https://iopscience.iop.org/article/10.3847/1538-4357/ab4f7a},
    doi = {10.3847/1538-4357/ab4f7a},
    language = {en},
    number = {1},
    urldate = {2025-01-06},
    journal = {The Astrophysical Journal},
    author = {{The SunPy Community} and Barnes, Will T. and Bobra, Monica G. and Christe, Steven D. and Freij, Nabil and Hayes, Laura A. and Ireland, Jack and Mumford, Stuart and Perez-Suarez, David and Ryan, Daniel F. and Shih, Albert Y. and {(Primary Paper Contributors)} and Chanda, Prateek and Glogowski, Kolja and Hewett, Russell and Hughitt, V. Keith and Hill, Andrew and Hiware, Kaustubh and Inglis, Andrew and Kirk, Michael S. F. and Konge, Sudarshan and Mason, James Paul and Maloney, Shane Anthony and Murray, Sophie A. and Panda, Asish and Park, Jongyeob and Pereira, Tiago M. D. and Reardon, Kevin and Savage, Sabrina and Sipőcz, Brigitta M. and Stansby, David and Jain, Yash and Taylor, Garrison and Yadav, Tannmay and {Rajul} and Dang, Trung Kien and {(Sunpy Contributors)}},
    month = feb,
    year = {2020},
    pages = {68},
}

@article{the_astropy_collaboration_astropy_2022,
    title = {The {Astropy} {Project}: {Sustaining} and {Growing} a {Community}-oriented {Open}-source {Project} and the {Latest} {Major} {Release} (v5.0) of the {Core} {Package}*},
    volume = {935},
    issn = {0004-637X, 1538-4357},
    shorttitle = {The {Astropy} {Project}},
    url = {https://iopscience.iop.org/article/10.3847/1538-4357/ac7c74},
    doi = {10.3847/1538-4357/ac7c74},
    language = {en},
    number = {2},
    urldate = {2025-04-18},
    journal = {The Astrophysical Journal},
    author = {{The Astropy Collaboration} and Price-Whelan, Adrian M. and Lim, Pey Lian and Earl, Nicholas and Starkman, Nathaniel and Bradley, Larry and Shupe, David L. and Patil, Aarya A. and Corrales, Lia and Brasseur, C. E. and Nöthe, Maximilian and Donath, Axel and Tollerud, Erik and Morris, Brett M. and Ginsburg, Adam and Vaher, Eero and Weaver, Benjamin A. and Tocknell, James and Jamieson, William and Van Kerkwijk, Marten H. and Robitaille, Thomas P. and Merry, Bruce and Bachetti, Matteo and Günther, H. Moritz and {Paper Authors} and Aldcroft, Thomas L. and Alvarado-Montes, Jaime A. and Archibald, Anne M. and Bódi, Attila and Bapat, Shreyas and Barentsen, Geert and Bazán, Juanjo and Biswas, Manish and Boquien, Médéric and Burke, D. J. and Cara, Daria and Cara, Mihai and Conroy, Kyle E and Conseil, Simon and Craig, Matthew W. and Cross, Robert M. and Cruz, Kelle L. and D’Eugenio, Francesco and Dencheva, Nadia and Devillepoix, Hadrien A. R. and Dietrich, Jörg P. and Eigenbrot, Arthur Davis and Erben, Thomas and Ferreira, Leonardo and Foreman-Mackey, Daniel and Fox, Ryan and Freij, Nabil and Garg, Suyog and Geda, Robel and Glattly, Lauren and Gondhalekar, Yash and Gordon, Karl D. and Grant, David and Greenfield, Perry and Groener, Austen M. and Guest, Steve and Gurovich, Sebastian and Handberg, Rasmus and Hart, Akeem and Hatfield-Dodds, Zac and Homeier, Derek and Hosseinzadeh, Griffin and Jenness, Tim and Jones, Craig K. and Joseph, Prajwel and Kalmbach, J. Bryce and Karamehmetoglu, Emir and Kałuszyński, Mikołaj and Kelley, Michael S. P. and Kern, Nicholas and Kerzendorf, Wolfgang E. and Koch, Eric W. and Kulumani, Shankar and Lee, Antony and Ly, Chun and Ma, Zhiyuan and MacBride, Conor and Maljaars, Jakob M. and Muna, Demitri and Murphy, N. A. and Norman, Henrik and O’Steen, Richard and Oman, Kyle A. and Pacifici, Camilla and Pascual, Sergio and Pascual-Granado, J. and Patil, Rohit R. and Perren, Gabriel I and Pickering, Timothy E. and Rastogi, Tanuj and Roulston, Benjamin R. and Ryan, Daniel F and Rykoff, Eli S. and Sabater, Jose and Sakurikar, Parikshit and Salgado, Jesús and Sanghi, Aniket and Saunders, Nicholas and Savchenko, Volodymyr and Schwardt, Ludwig and Seifert-Eckert, Michael and Shih, Albert Y. and Jain, Anany Shrey and Shukla, Gyanendra and Sick, Jonathan and Simpson, Chris and Singanamalla, Sudheesh and Singer, Leo P. and Singhal, Jaladh and Sinha, Manodeep and Sipőcz, Brigitta M. and Spitler, Lee R. and Stansby, David and Streicher, Ole and Šumak, Jani and Swinbank, John D. and Taranu, Dan S. and Tewary, Nikita and Tremblay, Grant R. and Val-Borro, Miguel De and Van Kooten, Samuel J. and Vasović, Zlatan and Verma, Shresth and De Miranda Cardoso, José Vinícius and Williams, Peter K. G. and Wilson, Tom J. and Winkel, Benjamin and Wood-Vasey, W. M. and Xue, Rui and Yoachim, Peter and Zhang, Chen and Zonca, Andrea and {Astropy Project Contributors}},
    month = aug,
    year = {2022},
    pages = {167},
}

@misc{markwardt_non-linear_2009,
    title = {Non-linear {Least} {Squares} {Fitting} in {IDL} with {MPFIT}},
    url = {http://arxiv.org/abs/0902.2850},
    doi = {10.48550/arXiv.0902.2850},
    language = {en},
    urldate = {2025-04-18},
    publisher = {arXiv},
    author = {Markwardt, Craig B.},
    month = feb,
    year = {2009},
    note = {arXiv:0902.2850 [astro-ph]},
}

@article{weberg_eispac_2023,
    title = {{EISPAC} - {The} {EIS} {Python} {Analysis} {Code}},
    volume = {8},
    copyright = {http://creativecommons.org/licenses/by/4.0/},
    issn = {2475-9066},
    url = {https://joss.theoj.org/papers/10.21105/joss.04914},
    doi = {10.21105/joss.04914},
    language = {en},
    number = {85},
    urldate = {2025-01-06},
    journal = {Journal of Open Source Software},
    author = {Weberg, Micah J. and Warren, Harry P. and Crump, Nicholas and Barnes, Will},
    month = may,
    year = {2023},
    pages = {4914},
}

@article{young_velocity_2012,
    title = {Velocity {Measurements} for a {Solar} {Active} {Region} {Fan} {Loop} from {Hinode}/{EIS} {Observations}},
    volume = {744},
    issn = {0004-637X, 1538-4357},
    url = {https://iopscience.iop.org/article/10.1088/0004-637X/744/1/14},
    doi = {10.1088/0004-637X/744/1/14},
    language = {en},
    number = {1},
    urldate = {2025-01-06},
    journal = {The Astrophysical Journal},
    author = {Young, P. R. and O’Dwyer, B. and Mason, H. E.},
    month = jan,
    year = {2012},
    pages = {14},
}

@article{young_euv_2007,
    title = {{EUV} {Emission} {Lines} and {Diagnostics} {Observed} with {Hinode}/{EIS}},
    volume = {59},
    issn = {2053-051X, 0004-6264},
    url = {https://academic.oup.com/pasj/article/59/sp3/S857/1406619},
    doi = {10.1093/pasj/59.sp3.S857},
    language = {en},
    number = {sp3},
    urldate = {2025-01-06},
    journal = {Publications of the Astronomical Society of Japan},
    author = {Young, Peter R. and Zanna, Del Giulio and Mason, Helen E. and Dere, Ken P. and Landi, Enrico and Landini, Massimo and Doschek, George A. and Brown, Charles M. and Culhane, Len and Harra, Louise K. and Watanabe, Tetsuya and Hara, Hirohisa},
    month = nov,
    year = {2007},
    pages = {S857--S864},
}

@article{del_zanna_benchmarking_2005,
    title = {Benchmarking atomic data for astrophysics: {Fe} {XII}},
    volume = {433},
    issn = {0004-6361, 1432-0746},
    shorttitle = {Benchmarking atomic data for astrophysics},
    url = {http://www.aanda.org/10.1051/0004-6361:20041848},
    doi = {10.1051/0004-6361:20041848},
    language = {en},
    number = {2},
    urldate = {2025-01-06},
    journal = {Astronomy \& Astrophysics},
    author = {Del Zanna, G. and Mason, H. E.},
    month = apr,
    year = {2005},
    pages = {731--744},
}

@article{demoulin_3d_2013,
    title = {The {3D} {Geometry} of {Active} {Region} {Upflows} {Deduced} from {Their} {Limb}-to-{Limb} {Evolution}},
    volume = {283},
    copyright = {http://www.springer.com/tdm},
    issn = {0038-0938, 1573-093X},
    url = {http://link.springer.com/10.1007/s11207-013-0234-7},
    doi = {10.1007/s11207-013-0234-7},
    language = {en},
    number = {2},
    urldate = {2025-04-24},
    journal = {Solar Physics},
    author = {Démoulin, P. and Baker, D. and Mandrini, C. H. and Van Driel-Gesztelyi, L.},
    month = apr,
    year = {2013},
    pages = {341--367},
}

@book{bevington_data_2003,
    address = {Boston, Mass.},
    edition = {3},
    title = {Data reduction and error analysis for the physical sciences},
    isbn = {978-0-07-247227-1 978-93-392-2120-1},
    language = {eng},
    publisher = {McGraw-Hill},
    author = {Bevington, Philip R. and Robinson, Keith D.},
    year = {2003},
}

@article{chae_sumer_1998,
    title = {{SUMER} {Measurements} of {Nonthermal} {Motions}: {Constraints} on {Coronal} {Heating} {Mechanisms}},
    volume = {505},
    issn = {0004-637X, 1538-4357},
    shorttitle = {{SUMER} {Measurements} of {Nonthermal} {Motions}},
    url = {https://iopscience.iop.org/article/10.1086/306179},
    doi = {10.1086/306179},
    language = {en},
    number = {2},
    urldate = {2025-01-06},
    journal = {The Astrophysical Journal},
    author = {Chae, Jongchul and Schuhle, Udo and Lemaire, Philippe},
    month = oct,
    year = {1998},
    pages = {957--973},
}

@article{watanabe_fe_2009,
    title = {Fe {XIII} {Density} {Diagnostics} in the {EIS} {Observing} {Wavelengths}},
    volume = {692},
    issn = {0004-637X, 1538-4357},
    url = {https://iopscience.iop.org/article/10.1088/0004-637X/692/2/1294},
    doi = {10.1088/0004-637X/692/2/1294},
    language = {en},
    number = {2},
    urldate = {2025-01-06},
    journal = {The Astrophysical Journal},
    author = {Watanabe, T. and Hara, H. and Yamamoto, N. and Kato, D. and Sakaue, H. A. and Murakami, I. and Kato, T. and Nakamura, N. and Young, P. R.},
    month = feb,
    year = {2009},
    pages = {1294--1304},
}

@article{warren_absolute_2014,
    title = {The {Absolute} {Calibration} of the {EUV} {Imaging} {Spectrometer} on {Hinode}},
    volume = {213},
    copyright = {http://iopscience.iop.org/info/page/text-and-data-mining},
    issn = {0067-0049, 1538-4365},
    url = {https://iopscience.iop.org/article/10.1088/0067-0049/213/1/11},
    doi = {10.1088/0067-0049/213/1/11},
    language = {en},
    number = {1},
    urldate = {2025-01-06},
    journal = {The Astrophysical Journal Supplement Series},
    author = {Warren, Harry P. and Ugarte-Urra, Ignacio and Landi, Enrico},
    month = jun,
    year = {2014},
    pages = {11},
}

@article{lemen_atmospheric_2012,
    title = {The {Atmospheric} {Imaging} {Assembly} ({AIA}) on the {Solar} {Dynamics} {Observatory} ({SDO})},
    volume = {275},
    issn = {0038-0938, 1573-093X},
    url = {https://link.springer.com/10.1007/s11207-011-9776-8},
    doi = {10.1007/s11207-011-9776-8},
    language = {en},
    number = {1-2},
    urldate = {2025-03-02},
    journal = {Solar Physics},
    author = {Lemen, James R. and Title, Alan M. and Akin, David J. and Boerner, Paul F. and Chou, Catherine and Drake, Jerry F. and Duncan, Dexter W. and Edwards, Christopher G. and Friedlaender, Frank M. and Heyman, Gary F. and Hurlburt, Neal E. and Katz, Noah L. and Kushner, Gary D. and Levay, Michael and Lindgren, Russell W. and Mathur, Dnyanesh P. and McFeaters, Edward L. and Mitchell, Sarah and Rehse, Roger A. and Schrijver, Carolus J. and Springer, Larry A. and Stern, Robert A. and Tarbell, Theodore D. and Wuelser, Jean-Pierre and Wolfson, C. Jacob and Yanari, Carl and Bookbinder, Jay A. and Cheimets, Peter N. and Caldwell, David and Deluca, Edward E. and Gates, Richard and Golub, Leon and Park, Sang and Podgorski, William A. and Bush, Rock I. and Scherrer, Philip H. and Gummin, Mark A. and Smith, Peter and Auker, Gary and Jerram, Paul and Pool, Peter and Soufli, Regina and Windt, David L. and Beardsley, Sarah and Clapp, Matthew and Lang, James and Waltham, Nicholas},
    month = jan,
    year = {2012},
    pages = {17--40},
}

@article{pesnell_solar_2012,
    title = {The {Solar} {Dynamics} {Observatory} ({SDO})},
    volume = {275},
    issn = {0038-0938, 1573-093X},
    url = {http://link.springer.com/10.1007/s11207-011-9841-3},
    doi = {10.1007/s11207-011-9841-3},
    language = {en},
    number = {1-2},
    urldate = {2025-03-02},
    journal = {Solar Physics},
    author = {Pesnell, W. Dean and Thompson, B. J. and Chamberlin, P. C.},
    month = jan,
    year = {2012},
    pages = {3--15},
}

@article{hoeksema_helioseismic_2014,
    title = {The {Helioseismic} and {Magnetic} {Imager} ({HMI}) {Vector} {Magnetic} {Field} {Pipeline}: {Overview} and {Performance}},
    volume = {289},
    copyright = {http://creativecommons.org/licenses/by/4.0},
    issn = {0038-0938, 1573-093X},
    shorttitle = {The {Helioseismic} and {Magnetic} {Imager} ({HMI}) {Vector} {Magnetic} {Field} {Pipeline}},
    url = {http://link.springer.com/10.1007/s11207-014-0516-8},
    doi = {10.1007/s11207-014-0516-8},
    language = {en},
    number = {9},
    urldate = {2025-01-06},
    journal = {Solar Physics},
    author = {Hoeksema, J. Todd and Liu, Yang and Hayashi, Keiji and Sun, Xudong and Schou, Jesper and Couvidat, Sebastien and Norton, Aimee and Bobra, Monica and Centeno, Rebecca and Leka, K. D. and Barnes, Graham and Turmon, Michael},
    month = sep,
    year = {2014},
    pages = {3483--3530},
}

@book{bravais_analyse_1846,
    address = {Paris},
    title = {Analyse {Mathematique}. {Sur} les probabilités des erreurs de situation d'un point},
    publisher = {Imprimerie Royale},
    author = {Bravais, Auguste},
    year = {1846},
}

@article{winebarger_can_2004,
    title = {Can {TRACE} {Extreme}-{Ultraviolet} {Observations} of {Cooling} {Coronal} {Loops} {Be} {Used} to {Determine} the {Heating} {Parameters}?},
    volume = {610},
    issn = {0004-637X, 1538-4357},
    url = {https://iopscience.iop.org/article/10.1086/423304},
    doi = {10.1086/423304},
    language = {en},
    number = {2},
    urldate = {2025-09-26},
    journal = {The Astrophysical Journal},
    author = {Winebarger, Amy R. and Warren, Harry P.},
    month = aug,
    year = {2004},
    pages = {L129--L132},
}

@article{martinez-sykora_impact_2023,
    title = {The {Impact} of {Multifluid} {Effects} in the {Solar} {Chromosphere} on the {Ponderomotive} {Force} under {SE} and {NEQ} {Ionization} {Conditions}},
    volume = {949},
    issn = {0004-637X},
    url = {https://ui.adsabs.harvard.edu/abs/2023ApJ...949..112M},
    doi = {10.3847/1538-4357/acc465},
    urldate = {2025-09-26},
    journal = {The Astrophysical Journal},
    author = {Martínez-Sykora, Juan and De Pontieu, Bart and Hansteen, Viggo H. and Testa, Paola and Wargnier, Q. M. and Szydlarski, Mikolaj},
    month = jun,
    year = {2023},
    note = {Publisher: IOP
ADS Bibcode: 2023ApJ...949..112M},
    pages = {112},
}

@article{laming_first_2017,
    title = {The {First} {Ionization} {Potential} {Effect} from the {Ponderomotive} {Force}: {On} the {Polarization} and {Coronal} {Origin} of {Alfvén} {Waves}},
    volume = {844},
    issn = {0004-637X},
    shorttitle = {The {First} {Ionization} {Potential} {Effect} from the {Ponderomotive} {Force}},
    url = {https://ui.adsabs.harvard.edu/abs/2017ApJ...844..153L},
    doi = {10.3847/1538-4357/aa7cf1},
    urldate = {2025-09-26},
    journal = {The Astrophysical Journal},
    author = {Laming, J. Martin},
    month = aug,
    year = {2017},
    note = {ADS Bibcode: 2017ApJ...844..153L},
    pages = {153},
}

@article{pevtsov_relationship_2003,
    title = {The {Relationship} {Between} {X}-{Ray} {Radiance} and {Magnetic} {Flux}},
    volume = {598},
    issn = {0004-637X},
    url = {https://ui.adsabs.harvard.edu/abs/2003ApJ...598.1387P},
    doi = {10.1086/378944},
    urldate = {2025-09-26},
    journal = {The Astrophysical Journal},
    author = {Pevtsov, Alexei A. and Fisher, George H. and Acton, Loren W. and Longcope, Dana W. and Johns-Krull, Christopher M. and Kankelborg, Charles C. and Metcalf, Thomas R.},
    month = dec,
    year = {2003},
    note = {ADS Bibcode: 2003ApJ...598.1387P},
    pages = {1387--1391},
}

@article{young_scattered_2022,
    title = {Scattered {Light} in the {Hinode}/{EIS} and {SDO}/{AIA} {Instruments} {Measured} from the 2012 {Venus} {Transit}},
    volume = {938},
    issn = {0004-637X, 1538-4357},
    url = {https://iopscience.iop.org/article/10.3847/1538-4357/ac8472},
    doi = {10.3847/1538-4357/ac8472},
    number = {1},
    urldate = {2025-09-26},
    journal = {The Astrophysical Journal},
    author = {Young, Peter R. and Viall, Nicholeen M.},
    month = oct,
    year = {2022},
    pages = {27},
}

@article{ugarte-urra_magnetic_2015,
    title = {Magnetic {Flux} {Transport} and the {Long}-term {Evolution} of {Solar} {Active} {Regions}},
    volume = {815},
    issn = {0004-637X},
    url = {https://ui.adsabs.harvard.edu/abs/2015ApJ...815...90U},
    doi = {10.1088/0004-637X/815/2/90},
    urldate = {2025-09-26},
    journal = {The Astrophysical Journal},
    author = {Ugarte-Urra, Ignacio and Upton, Lisa and Warren, Harry P. and Hathaway, David H.},
    month = dec,
    year = {2015},
    note = {ADS Bibcode: 2015ApJ...815...90U},
    pages = {90},
}

@techreport{young_instrumental_2011,
    address = {Washington, DC},
    type = {Technical {Report}},
    title = {Instrumental line widths for the narrow slits of {EIS}},
    url = {https://solarb.mssl.ucl.ac.uk/SolarB/eis_docs/eis_notes/07_LINE_WIDTH/eis_swnote_07.pdf},
    language = {en},
    number = {7 (1.0)},
    institution = {Naval Research Laboratory},
    author = {Young, Peter},
    month = oct,
    year = {2011},
}

@article{del_zanna_hinode_2025,
    title = {Hinode {EIS}: {Updated} {In}-flight {Radiometric} {Calibration}},
    volume = {276},
    issn = {0067-0049, 1538-4365},
    shorttitle = {Hinode {EIS}},
    url = {https://iopscience.iop.org/article/10.3847/1538-4365/ad981f},
    doi = {10.3847/1538-4365/ad981f},
    number = {2},
    urldate = {2025-09-27},
    journal = {The Astrophysical Journal Supplement Series},
    author = {Del Zanna, G. and Weberg, M. J. and Warren, H. P.},
    month = feb,
    year = {2025},
    pages = {42},
}

@article{scherrer_helioseismic_2012,
    title = {The {Helioseismic} and {Magnetic} {Imager} ({HMI}) {Investigation} for the {Solar} {Dynamics} {Observatory} ({SDO})},
    volume = {275},
    issn = {0038-0938},
    url = {https://ui.adsabs.harvard.edu/abs/2012SoPh..275..207S},
    doi = {10.1007/s11207-011-9834-2},
    urldate = {2025-10-06},
    journal = {Solar Physics},
    author = {Scherrer, P. H. and Schou, J. and Bush, R. I. and Kosovichev, A. G. and Bogart, R. S. and Hoeksema, J. T. and Liu, Y. and Duvall, T. L. and Zhao, J. and Title, A. M. and Schrijver, C. J. and Tarbell, T. D. and Tomczyk, S.},
    month = jan,
    year = {2012},
    note = {ADS Bibcode: 2012SoPh..275..207S},
    pages = {207--227},
}

@article{alfven_magneto_1947,
    title = {Magneto hydrodynamic waves, and the heating of the solar corona},
    volume = {107},
    issn = {0035-8711},
    url = {https://ui.adsabs.harvard.edu/abs/1947MNRAS.107..211A},
    doi = {10.1093/mnras/107.2.211},
    urldate = {2025-11-18},
    journal = {Monthly Notices of the Royal Astronomical Society},
    author = {Alfvén, H.},
    month = jan,
    year = {1947},
    note = {Publisher: OUP
ADS Bibcode: 1947MNRAS.107..211A},
    pages = {211},
}

@article{ionson_resonant_1978,
    title = {Resonant absorption of {Alfvénic} surface waves and the heating of solar coronal loops.},
    volume = {226},
    issn = {0004-637X},
    url = {https://ui.adsabs.harvard.edu/abs/1978ApJ...226..650I},
    doi = {10.1086/156648},
    urldate = {2025-11-18},
    journal = {The Astrophysical Journal},
    author = {Ionson, J. A.},
    month = dec,
    year = {1978},
    note = {Publisher: IOP
ADS Bibcode: 1978ApJ...226..650I},
    pages = {650--673},
}

@article{fan_origin_1993,
    title = {The {Origin} of {Morphological} {Asymmetries} in {Bipolar} {Active} {Regions}},
    volume = {405},
    issn = {0004-637X},
    url = {https://ui.adsabs.harvard.edu/abs/1993ApJ...405..390F},
    doi = {10.1086/172370},
    urldate = {2026-01-14},
    journal = {The Astrophysical Journal},
    publisher = {IOP},
    author = {Fan, Y. and Fisher, G. H. and Deluca, E. E.},
    month = mar,
    year = {1993},
    note = {ADS Bibcode: 1993ApJ...405..390F},
    pages = {390},
}

@article{kamio_modeling_2010,
    title = {Modeling of {EIS} {Spectrum} {Drift} from {Instrumental} {Temperatures}},
    volume = {266},
    copyright = {http://www.springer.com/tdm},
    issn = {0038-0938, 1573-093X},
    url = {http://link.springer.com/10.1007/s11207-010-9603-7},
    doi = {10.1007/s11207-010-9603-7},
    language = {en},
    number = {1},
    urldate = {2025-08-21},
    journal = {Solar Physics},
    author = {Kamio, S. and Hara, H. and Watanabe, T. and Fredvik, T. and Hansteen, V. H.},
    month = sep,
    year = {2010},
    pages = {209--223},
}

@article{morton_alfvenic_2023,
    title = {Alfvénic waves in the inhomogeneous solar atmosphere},
    volume = {7},
    issn = {2367-3192},
    url = {https://link.springer.com/10.1007/s41614-023-00118-3},
    doi = {10.1007/s41614-023-00118-3},
    language = {en},
    number = {1},
    urldate = {2026-01-26},
    journal = {Reviews of Modern Plasma Physics},
    author = {Morton, R. J. and Sharma, R. and Tajfirouze, E. and Miriyala, H.},
    month = mar,
    year = {2023},
    pages = {17},
}

@article{mckevitt_pre-flare_2026,
    title = {Pre-flare and active region plasma motion seen by the short wavelength camera on {SOLAR}-{C}/{EUVST}},
    journal = {submitted},
    author = {McKevitt, James and Matthews, Sarah and Brooks, David and Shimizu, Toshifumi and Tei, Akiko and Ugarte-Urra, Ignacio and Imada, Shinsuke and Brown, Charles and Ishikawa, Ryohko and Katsukawa, Yukio and Rust, Duncan and Walton, Dave and Winter, Berend and Baker, Deborah and Reid, Hamish and Young, Peter and Pereira, Tiago and Bradley, Louisa and Shitvov, Alexey and Harra, Louise and {International SOLAR-C team}},
    year = {2026},
}

\end{document}